\documentclass[]{emulateapj}

\usepackage{natbib}
\usepackage{natbib}
\usepackage{amsmath}
\usepackage{graphicx}
\usepackage{hyperref}

\def\nicefrac#1#2{
    \raise.5ex\hbox{#1}%
    \kern-.1em/\kern-.15em%
    \lower.25ex\hbox{#2}}

\begin{document}



\title{LOFAR facet calibration}
\author{R.~J.~van~Weeren\altaffilmark{1,$\star$}, W.~L.~Williams\altaffilmark{2,3,4}, M.~J.~Hardcastle\altaffilmark{4}, T.~W.~Shimwell\altaffilmark{2}, D.~A.~Rafferty\altaffilmark{5},  J.~Sabater\altaffilmark{6}, G.~Heald\altaffilmark{3,7}, S.~S.~{Sridhar}\altaffilmark{7,3}, T.~J.~Dijkema\altaffilmark{3},  
G.~Brunetti\altaffilmark{8},  M.~Br\"uggen\altaffilmark{5}, F.~Andrade-Santos\altaffilmark{1}, G.~A.~Ogrean\altaffilmark{1,$\dagger$}, H.~J.~A.~R\"ottgering\altaffilmark{2}, W.~A.~Dawson\altaffilmark{9}, W.~R.~Forman\altaffilmark{1},   F.~de~Gasperin\altaffilmark{2,5}, C.~Jones\altaffilmark{1}, G.~K.~Miley\altaffilmark{2}, L.~Rudnick\altaffilmark{10}, C.~L.~Sarazin\altaffilmark{11},  
A.~Bonafede\altaffilmark{5}, P.~N.~Best\altaffilmark{6}, L.~B{\^i}rzan\altaffilmark{5}, R.~Cassano\altaffilmark{8}, K.~T. Chy\.zy\altaffilmark{12}, J.~H.~Croston\altaffilmark{13},  T.~En{\ss}lin\altaffilmark{14}, C.~Ferrari\altaffilmark{15},   M.~Hoeft\altaffilmark{16}, C.~Horellou\altaffilmark{17}, M.~J.~Jarvis\altaffilmark{18,19}, R.~P.~Kraft\altaffilmark{1}, M.~Mevius\altaffilmark{3},  H.~T.~Intema\altaffilmark{20,2}, S.~S.~Murray\altaffilmark{1,21},  E.~Orr\'u\altaffilmark{3}, R.~Pizzo\altaffilmark{3}, A.~Simionescu\altaffilmark{22}, A.~Stroe\altaffilmark{2}, S. van der Tol\altaffilmark{3}, and G.~J.~White\altaffilmark{23,24}\vspace{3mm}
}

\affil{\altaffilmark{1}Harvard-Smithsonian Center for Astrophysics, 60 Garden Street, Cambridge, MA 02138, USA}
\affil{\altaffilmark{2}Leiden Observatory, Leiden University, P.O. Box 9513, NL-2300 RA Leiden, The Netherlands}
\affil{\altaffilmark{3}ASTRON, the Netherlands Institute for Radio Astronomy, Postbus 2, 7990 AA, Dwingeloo, The Netherlands}
\affil{\altaffilmark{4}School of Physics, Astronomy and Mathematics, University of Hertfordshire, College Lane, Hatfield AL10 9AB, UK}
\affil{\altaffilmark{5}Hamburger Sternwarte, Gojenbergsweg 112, 21029 Hamburg, Germany}
\affil{\altaffilmark{6}Institute for Astronomy, University of Edinburgh, Royal Observatory, Blackford Hill, Edinburgh EH9 3HJ, UK}
\affil{\altaffilmark{7}Kapteyn Astronomical Institute, P.O. Box 800, 9700 AV Groningen, The Netherlands}
\affil{\altaffilmark{8}INAF/Istituto di Radioastronomia, via Gobetti 101, I-40129 Bologna, Italy}
\affil{\altaffilmark{9}Lawrence Livermore National Lab, 7000 East Avenue, Livermore, CA 94550, USA}
\affil{\altaffilmark{10}Minnesota Institute for Astrophysics, University of Minnesota, 116 Church St. S.E., Minneapolis, MN 55455, USA}
\affil{\altaffilmark{11}Department of Astronomy, University of Virginia, Charlottesville, VA 22904-4325, USA}
\affil{\altaffilmark{12}Astronomical Observatory, Jagiellonian University, ul. Orla 171, 30-244 Krak\'ow, Poland}
\affil{\altaffilmark{13}School of Physics and Astronomy, University of Southampton, Southampton SO17 1BJ, UK}
\affil{\altaffilmark{14}Max Planck Institute for Astrophysics, Karl-Schwarzschildstr. 1, 85741 Garching, Germany}
\affil{\altaffilmark{15}Laboratoire Lagrange, Universit\'e C\^{o}te d'Azur, Observatoire de la C\^{o}te d'Azur, CNRS, Blvd de l'Observatoire, CS 34229, 06304 Nice cedex 4, France}
\affil{\altaffilmark{16}Th\"uringer Landessternwarte Tautenburg, Sternwarte 5, 07778, Tautenburg, Germany}
\affil{\altaffilmark{17}Department of Earth and Space Sciences, Chalmers University of Technology, Onsala Space Observatory, SE-43992 Onsala, Sweden} 
\affil{\altaffilmark{18}Oxford Astrophysics, Department of Physics, Keble Road, Oxford, OX1 3RH, UK}
\affil{\altaffilmark{19}University of the Western Cape, Bellville 7535, South Africa}
\affil{\altaffilmark{20}National Radio Astronomy Observatory, 1003 Lopezville Road, Socorro, NM 87801-0387, USA}
\affil{\altaffilmark{21}Department of Physics and Astronomy, Johns Hopkins University, 3400 North Charles Street, Baltimore, MD 21218, USA}
\affil{\altaffilmark{22}Institute of Space and Astronautical Science (ISAS), JAXA, 3-1-1 Yoshinodai, Chuo, Sagamihara, Kanagawa, 252-5210, Japan}
\affil{\altaffilmark{23}Department of Physical Sciences, The Open University, Walton Hall, Milton Keynes MK7 6AA, UK}
\affil{\altaffilmark{24}RALSpace, The Rutherford Appleton Laboratory, Chilton, Didcot, Oxfordshire OX11 0NL, UK}

\email{E-mail: rvanweeren@cfa.harvard.edu}

\altaffiltext{$\star$}{Einstein Fellow}
\altaffiltext{$\dagger$}{Hubble Fellow}

\shorttitle{LOFAR facet calibration}
\shortauthors{van Weeren et al.}

\vspace{0.5cm}
\begin{abstract}
\noindent

{LOFAR, the Low-Frequency Array, is a powerful new radio telescope operating between 10 and 240~MHz. LOFAR allows detailed sensitive high-resolution studies of the low-frequency radio sky. At the same time LOFAR also provides excellent short baseline coverage to map diffuse extended emission. However, producing high-quality deep images is challenging due to the presence of direction dependent calibration errors, caused by imperfect knowledge of the station beam shapes and the ionosphere. Furthermore, the large data volume and presence of station clock errors present additional difficulties. 
In this paper we present a new calibration scheme, which we name facet calibration, to obtain deep high-resolution LOFAR High Band Antenna images using the Dutch part of the array. This scheme solves and corrects the direction dependent errors in a number of facets that cover the observed field of view. Facet calibration provides close to thermal noise limited images for a typical 8~hr observing run at $\sim 5\arcsec$~resolution, meeting the specifications of the LOFAR Tier-1 northern survey.}

\vspace{4mm}
\end{abstract}

\section{Introduction}
\label{sec:intro}

{
LOFAR is a powerful new radio telescope operating between 10 to 240~MHz \citep{2013A&A...556A...2V}. It is designed to carry out a range of astrophysical studies in this relatively unexplored part of the radio band. LOFAR consists of  antenna dipoles that are grouped into stations. LOFAR stations are located in various countries in Europe, with the majority being in the north-east of the Netherlands. 

Two different dipole antenna types are used. The High Band Antennas (HBAs) cover the 110-240~MHz range and the Low Band Antennas (LBAs) the 10--90~MHz range.
At the stations, the dipoles signals are combined digitally into a phased array. The signals from these stations are then sent via high-speed fiber to a central GPU correlator where they are correlated with those from other stations to form an interferometer. The electronic beam-forming at the station level allows the generation of multiple beams on the sky  which, together with the large field of view (FoV) at these low-frequencies, makes LOFAR an ideal survey instrument. For more details about the instrument we refer the reader to the overview paper by \cite{2013A&A...556A...2V}.

One of the major goals of LOFAR is to carry out a survey of the northern sky as part of the LOFAR Surveys Key Science Project (Surveys KSP) \citep{2006astro.ph.10596RX}. For the HBA part of the survey, the aim is to reach a depth of $\sim 0.1$~mJy~beam$^{-1}$ at a resolution of $\sim 5\arcsec$~\citep[i.e., the Tier-1 survey depth,][]{2011JApA...32..557R} across the entire northern sky. Over the last years it has become clear that, to reach this depth, advanced calibration and processing techniques are needed. 

One of the main challenges for the calibration is the ionosphere {\citep[e.g.,][]{2005ASPC..345..399L,2009A&A...501.1185I}}. The ionosphere results in delay differences between antenna stations causing errors in the phases of the measured visibilities. The amount of phase change is directly related to the free electron column density along a line of sight through the ionosphere and the observing frequency. These ionospheric phase errors thus change with the viewing direction, and if the array is large, as is the case for LOFAR, differ from station to station. The phase errors result in shifting, deformation, and splitting of sources in the image plane, which if not corrected, cause deconvolution artifacts and an increase in the overall image noise.

Another challenge concerns the complex time-varying station beam shapes. The reason for the time-variation is that the stations have no moving parts. Sources are tracked by adjusting the delays between the dipole elements when the sources move across the sky. In addition, small differences in station beam models and the actual station beam shapes cause errors that need to be corrected for in order to produce high-quality images, in particular for bright sources.

In this work we present a new calibration scheme, facet calibration, that has enabled us to make images that reach the LOFAR HBA Tier-1 survey depth and resolution. Facet calibration was developed to process a LOFAR HBA observation of the ``Toothbrush'' galaxy cluster and we will use this particular dataset to lay out the method. The scientific results on this galaxy cluster are presented in \cite{vanweerenscience}. 

\{Facet calibration builds upon the ``peeling'' technique \citep[e.g.,][]{2004SPIE.5489..817N}, where calibration solutions in a discrete number of directions are obtained, similar to other low-frequency calibration schemes \citep[e.g., SPAM, Sagecal, and MeqTrees;][]{2009A&A...501.1185I,2011MNRAS.414.1656K,2011A&A...527A.107S,2010A&A...524A..61N}.}  


The layout of this paper is as follows. We start with a description of the observations and general characteristics of a typical LOFAR dataset in Section~\ref{sec:obs}. We provide an overview of the direction independent calibration and processing in Section \ref{sec:non-directional}. The corrections for direction dependent effects (DDE) are described in Section~\ref{sec:DDE}. We end with a discussion and conclusions in Sections~\ref{sec:dicussions} and \ref{sec:conclusions}.

\section{LOFAR HBA observations}
\label{sec:obs}

Below we describe the setup of the Cycle~0 Toothbrush cluster observations which will serve as a reference dataset to demonstrate the facet calibration scheme.

The Toothbrush cluster was observed on Feb 24, 2013, mostly during nighttime with the LOFAR High Band Antenna (HBA) stations. Two station beams were formed: one on the target and one on the nearby (8.3\degr~separation) calibrator 3C147.  By default, all four correlation products were recorded and the frequency band was divided into subbands, each 195.3125~kHz wide. Each subband was further divided into 64 channels. The integration time was set to 1~s to facilitate the removal of radio frequency interference (RFI). Complete frequency coverage between 112--181 MHz was obtained on the Toothbrush field, while the calibrator 3C147 was covered with 121 subbands\footnote{This number is constrained by the total amount of bandwidth that can be handled by the system.}, randomly spread between 112--181 MHz. An overview of the observations is given in Table~\ref{tab:observations}. 

\begin{figure*}
\begin{center}
\includegraphics[angle =0, trim =0cm 0cm 0cm 0cm,width=0.28\textwidth]{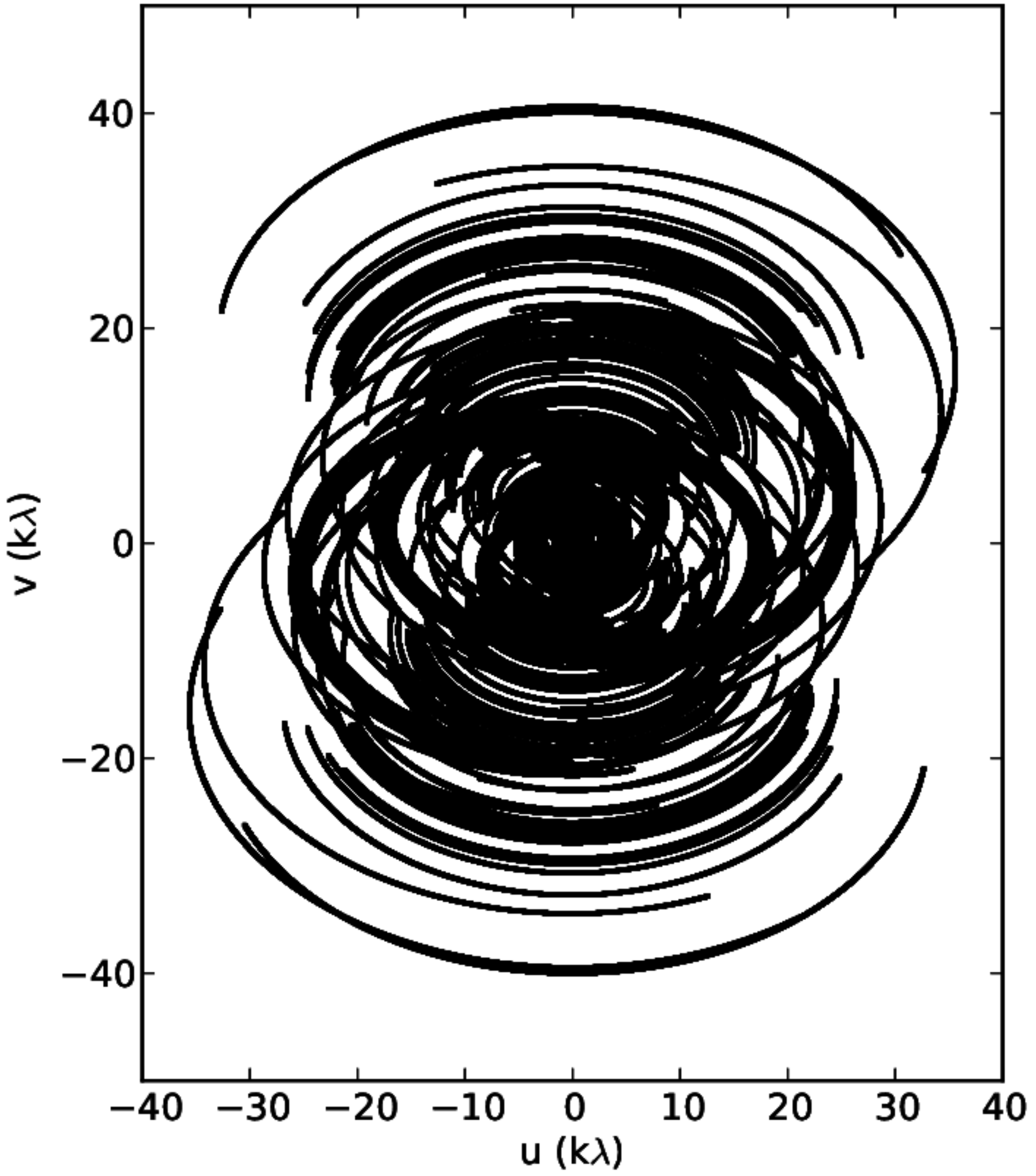}
\includegraphics[angle =0, trim =0cm 0cm 0cm 0cm,width=0.327\textwidth]{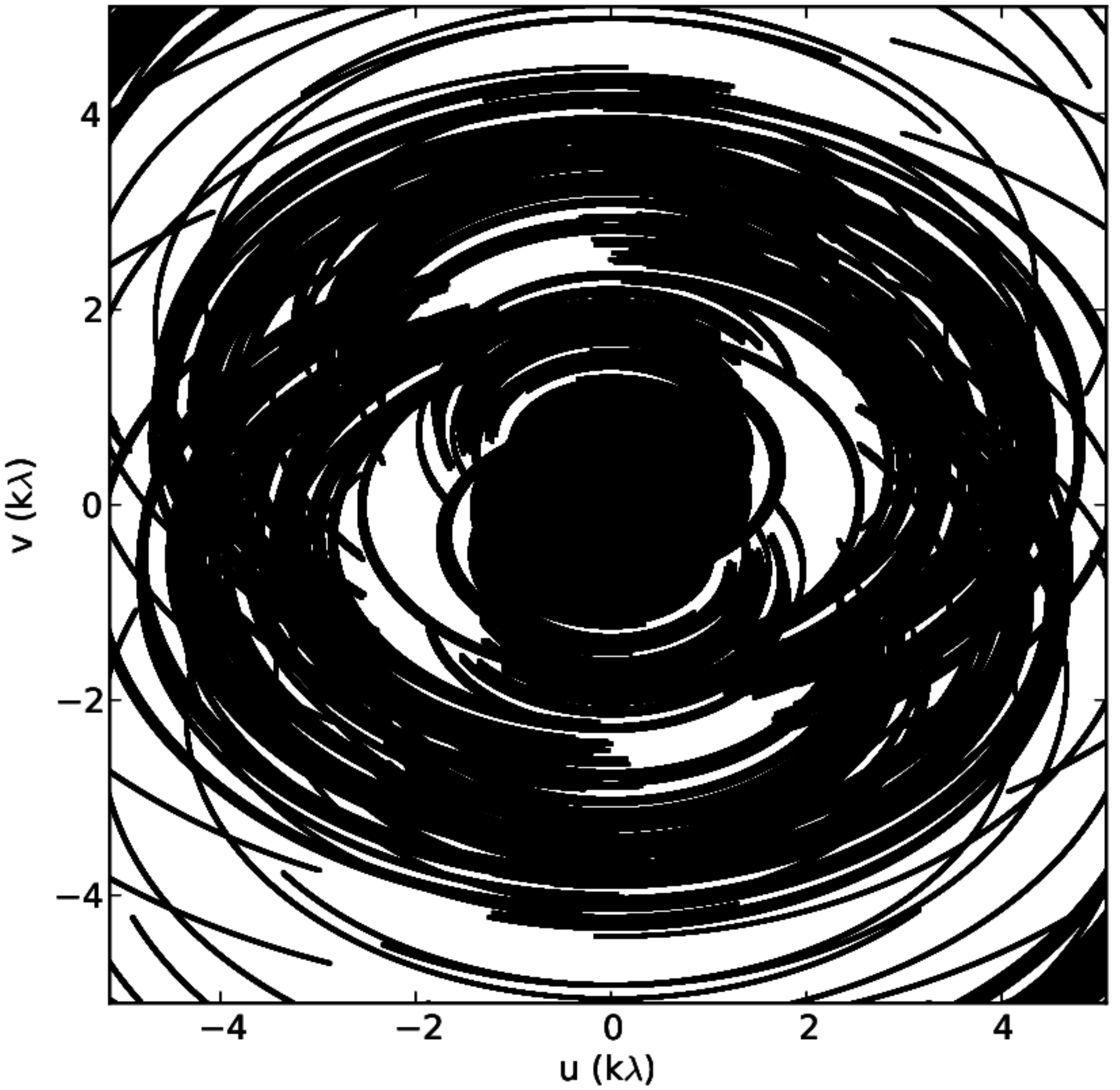}
\includegraphics[angle =0, trim =0cm 0cm 0cm 0cm,width=0.34\textwidth]{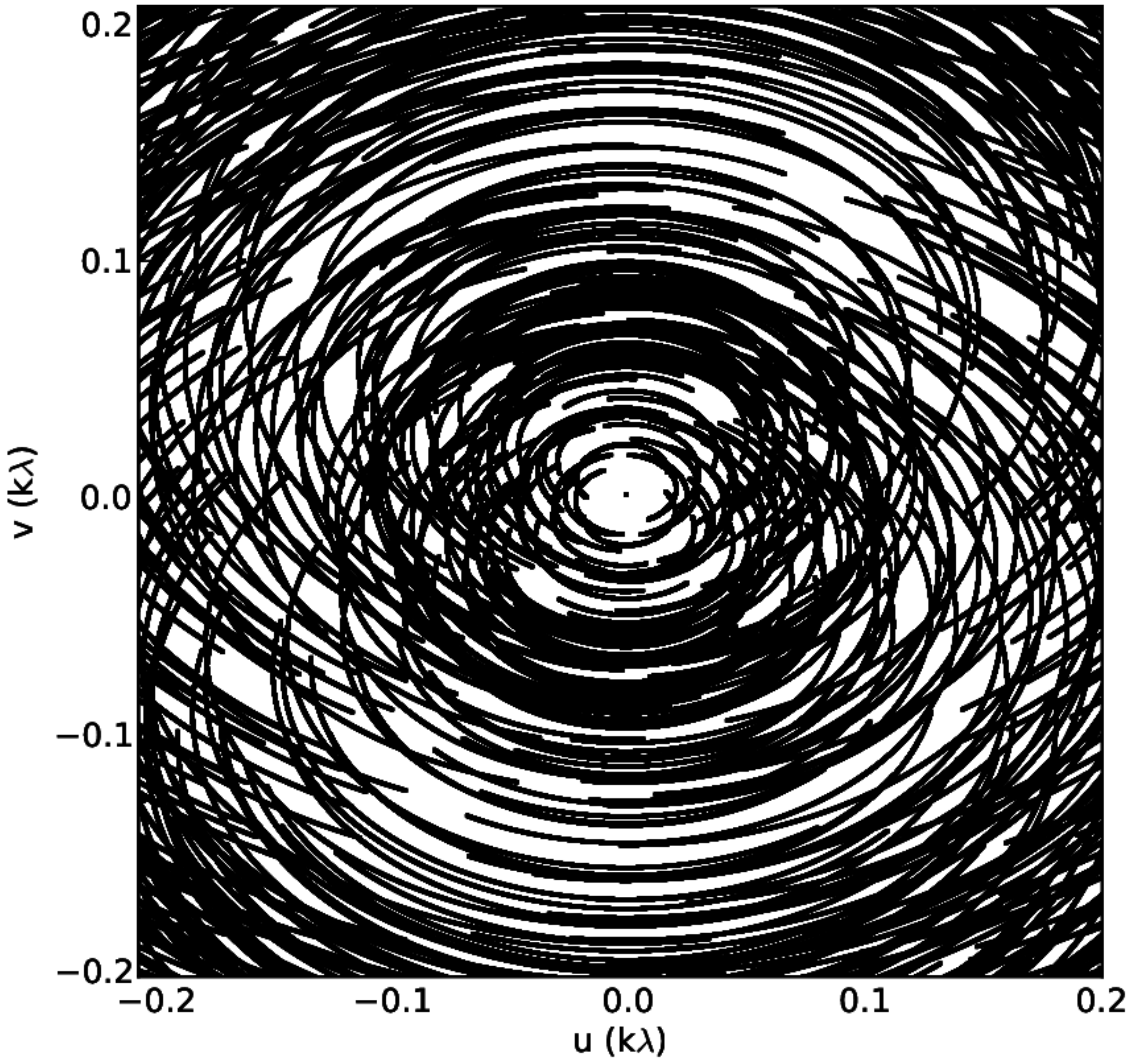}
\end{center}
\caption{Monochromatic uv-coverage for the Toothbrush field at 150~MHz. {The middle and right panels progressively zoom inwards showing the dense inner uv-coverage.}  The large fractional bandwidth fills the uv-plane radially (not shown in the figure). }
\label{fig:uv}
\end{figure*}

{For the observations 13 remote and 21 Dutch core stations were used, giving baselines that range between 68~m and 80~km. The large number of short baselines  is important to image diffuse extended emission.  The core station layout is different from that of the remote stations as the latter are split into two sub-stations, each with 24 dipole tiles and a diameter of 30.75~m. The HBA\_DUAL\_INNER configuration was employed for the remote stations \citep{2013A&A...556A...2V} meaning that only the inner 24 tiles are used of a remote station.  This is done to obtain similar station beam sizes as for the split core stations.  Thus, in total $21\times2 +13 =55$ stations were correlated for this observation. The international (``non-Dutch'') stations were not included.  For the stations, the half-power beam width (HPBW) is about 3.8\degr~at 150~MHz. The monochromatic uv-coverage for the Toothbrush field observations is displayed in Figure~\ref{fig:uv}.}

\begin{table}
\begin{center}
\caption{HBA Observations}
\begin{tabular}{lllll}
\hline
\hline
\hline
Observation IDs   & L99083, L99084 \\
Pointing center$^{a}$    & 06$^{\rm{h}}$03$^{\rm{m}}$33.5$^{\rm{s}}$, +42\degr19{\arcmin}58.5\arcsec  \\
Pointing center$^{b}$  & 05$^{\rm{h}}$42$^{\rm{m}}$36.1$^{\rm{s}}$, +49\degr51{\arcmin}07.0\arcsec  \\ 
Integration time     & 1~s \\
Observation date & 24 Feb, 2013 \\ 
Total on-source time		&  10~hr\\
Used on-source time	&  8.8~hr\\
Correlations           & XX, XY, YX, YY \\
Frequency setup$^{a}$               & 112--181~MHz full coverage\\
Frequency setup$^{b}$               & 112--181~MHz,  121 subbands$^{c}$\\ 
Used bandwidth on target & 120--181 MHz,  \\
 & except 169--171, 177--179~MHz$^{d}$ \\
Bandwidth per subband		  & 195.3125~kHz \\
Channels per subband                & 64\\
\hline
\hline
\end{tabular}
\label{tab:observations}
\end{center}
$^{a}$ Toothbrush field\\
$^{b}$ 3C147\\
$^{c}$ The subbands are approximately evenly distributed within this frequency range\\
$^{d}$ These frequency ranges are affected by strong RFI
\end{table}

\section{HBA data reduction and calibration: An overview}
\label{sec:hbadatareduction}
As mentioned in Section~\ref{sec:intro}, advanced calibration and processing techniques are needed to obtain deep high-fidelity images at low radio frequencies. 

 The Toothbrush observation was the first Cycle~0 observation of the LOFAR Galaxy Cluster Working group, which is part of the LOFAR surveys Key Science Project \citep{2006astro.ph.10596RX}. Therefore, the observation also served as a testbed to develop calibration strategies to reach the required Tier-1 survey depth. Below we provide an brief overview of the calibration method that was developed to reach the required depth and resolution, before we go into more detail in the next Sections.

The data reduction and calibration consists of two main components: a non-directional and a directional part. The non-directional part includes the following  steps: (1) removal of RFI, (2) bright off-axis source removal, (3) averaging, (4) solving for the calibrator complex gains, 
(5) ``clock-total electron content'' (clock-TEC) separation on the calibrator, (6) transfer of the amplitudes and clocks from the calibrator to the target field, and (7) amplitude and phase (self)calibration of the target field at  medium (20--30\arcsec)~resolution. This is then followed by a scheme to obtain direction dependent corrections to reach near thermal noise limited images using the full resolution offered by the longest ``Dutch-LOFAR'' baselines of about $10^2$~km. Below the non-directional and directional parts of the calibration are described in more detail.  

\section{Non-directional reduction and calibration}
\label{sec:non-directional}
\subsection{RFI removal}
The first step in the reduction of the HBA data consisted of removal of RFI with the {\tt AOFlagger} \citep{2010MNRAS.405..155O, 2012A&A...539A..95O}. The amount of data affected by RFI was typically only a few percent. One malfunctioning core station was flagged entirely. In addition, the first and last three channels of each subband were also flagged as they were noisy. After flagging, the data were averaged to 5~s and 4~channels per subband to reduce the data size.

\subsection{Removal of sources in the far sidelobes}
\label{sec:ateam}
 Cas~A and Cyg~A are sufficiently bright that they can contribute flux through the sidelobes of the station beam. We therefore computed the contribution of Cas~A and Cyg~A using the station beam model {\citep{hamaker,2013A&A...556A...2V}}. Whenever the apparent flux density of these sources exceeded 0.5~Jy for the core-core baselines, we obtained gain solutions towards these two sources using the {\tt BlackBoard Selfcal} ({\tt BBS}) software \citep{2009ASPC..407..384P}. In addition, simultaneous gain solutions were obtained towards 3C147 or the Toothbrush Field. We solved for the gains using all four correlations (XX, XY, YX, YY). For 3C147 we took a point source model. For the Toothbrush Field the model was derived from a GMRT 150~MHz image \citep[presented in][]{2012A&A...546A.124V} with the {\tt PyBDSM} \citep{2015ascl.soft02007M} source detection package.  The Cas~A and Cyg~A models came from $\sim10\arcsec$~resolution LOFAR low-band antenna observations. We assumed that all sources are unpolarized. After solving for the gains, Cas~A and Cyg~A were subtracted from the data with the appropriate gain solutions. The data were then averaged to 10~s and 2~channels {(97.6562~kHz per channel)} per subband for the target field and 5~s and 1~channel {(195.3125~kHz)} for 3C147; no gain corrections were applied. For 3C147, bandwidth smearing is not an issue as it is located in the phase center and dominates the flux in the field.  These data serve as input for the rest of the processing. The target field averaging parameters are a compromise between bandwidth and time smearing and finite computing capabilities.

\subsection{Obtaining calibration solutions towards the primary calibrator}
We obtained diagonal (XX and YY) gain solutions towards 3C147 with {\tt BBS},  solving on a timescale of 5~s per subband basis. Besides solving for these parallel-hand gains, we also solved for a  Rotation Angle ($\beta$) per station to take into account differential Faraday Rotation, which could otherwise affect the parallel-hand amplitudes in the linear correlation basis. The corresponding Jones matrix ($F$) for the differential Faraday Rotation \citep{2011A&A...527A.107S} is
\begin{eqnarray}
F =  \begin{pmatrix} \cos{\beta} & \mbox{ }-\sin{\beta} \\ \sin{\beta} & \cos{\beta} \end{pmatrix}  \mbox{ .}
\end{eqnarray}
We assume that $\beta$ is constant within a single subband, which is a valid approximation in the HBA frequency range. The  derived Rotation Angles  were small ($\beta \ll 1$~rad), indicating little differential Faraday Rotation, which is not unexpected in the HBA frequency range. The LOFAR station beam was applied during the solve so that gain variations caused by the changing station beams are not absorbed into the gain solutions. For the point source model of 3C147 we used the flux-scale of \cite{2012MNRAS.423L..30S},  giving a flux density of 66.7~Jy at 150~MHz.

\subsection{Clock-TEC separation}
\begin{figure*}
\begin{center}
\includegraphics[angle =0, trim =0cm 0cm 0cm 0cm,width=0.49\textwidth]{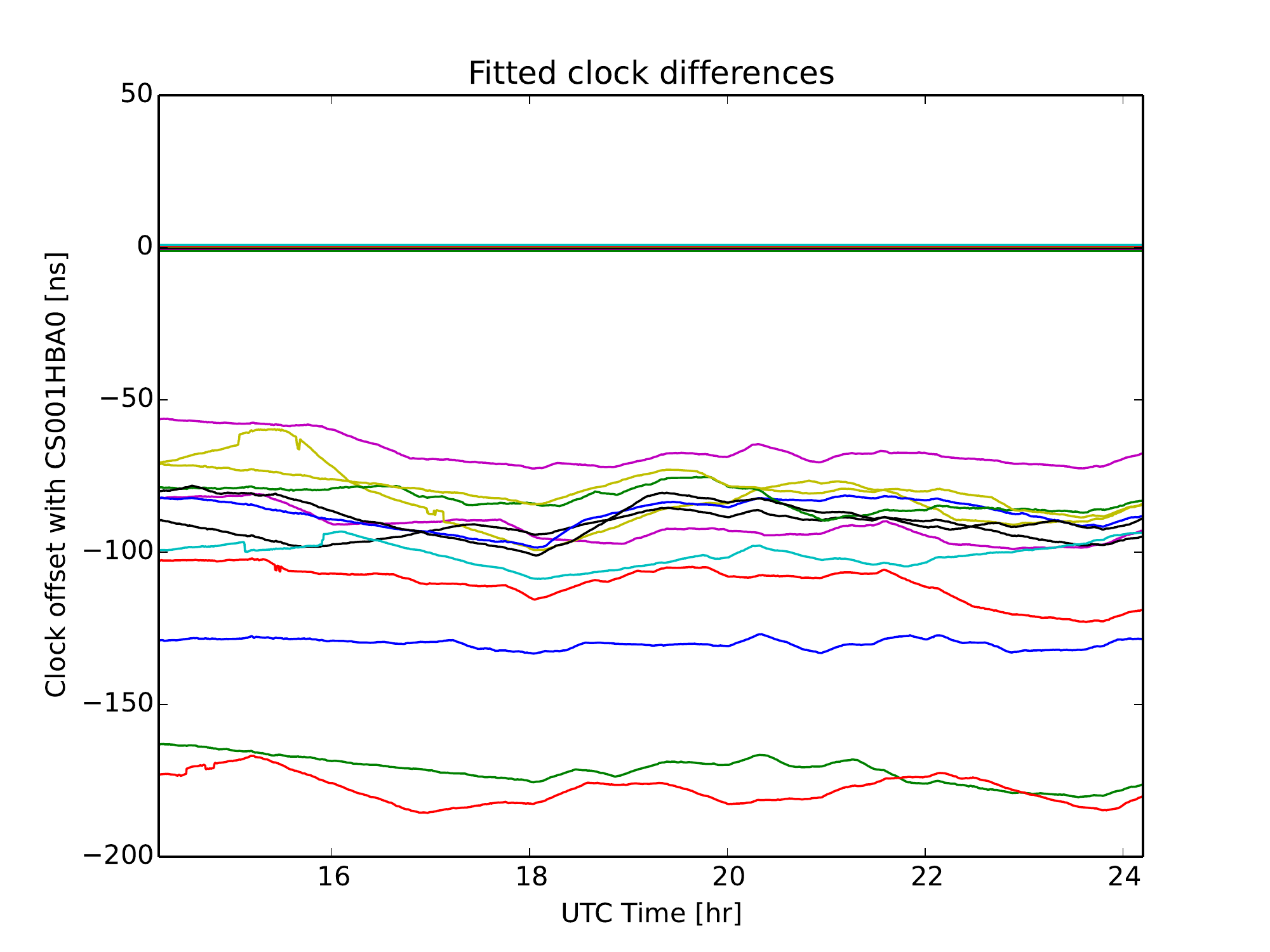}
\includegraphics[angle =0, trim =0cm 0cm 0cm 0cm,width=0.49\textwidth]{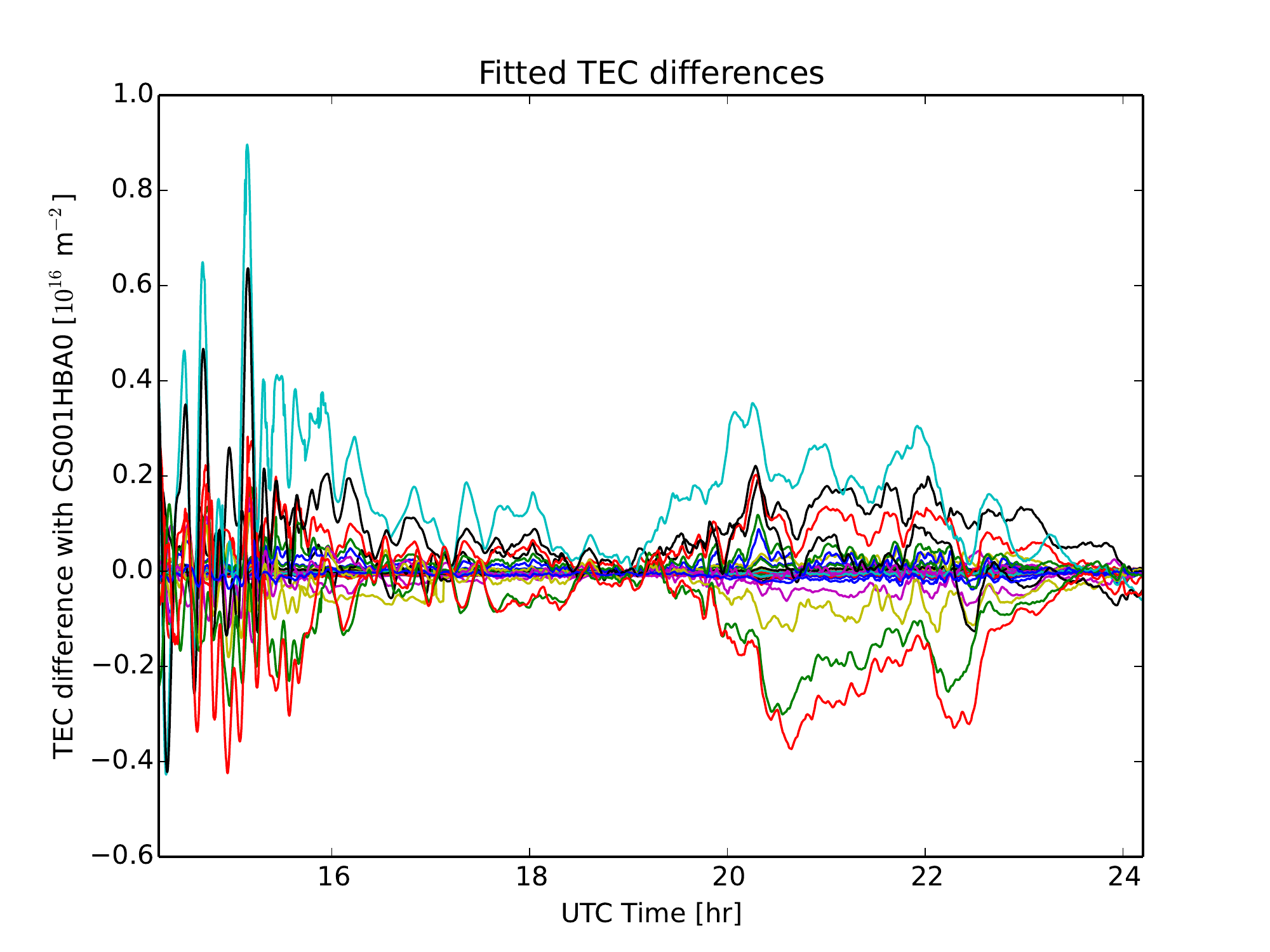}
\end{center}
\caption{Fitted clock and TEC differences based on the 3C147 phase solutions. The fitting was performed on the solution interval timescale of 5~s. The resulting values were smoothed with a running median filter with a local window size of 15~s. Left: Fitted station clock offsets as function of time with respect to the reference (core) station CS001HBA0. The clock values show a bimodal distribution with all the core stations having clock values close to zero, as is expected, since the core stations operate on a single clock \citep{2013A&A...556A...2V}. The remote stations all have large negative clock offsets with respect to the core stations, with the largest offset approaching 200~ns. {The fact that all remote stations have negative offsets indicates that it is the core station clock that has quite a large offset with respect to the average remote station clocks.} The changes in the slope of the clock drifts for the remote stations clocks are caused by adjustments of the clock rate. The clock rate adjustments are based on a comparison with a Global Positioning System (GPS) signal. These adjustments are made to prevent the clocks from drifting more than $\sim \pm15$~ns. Right: TEC values for the same stations. Differences of up to 1 TEC unit are observed which indicates a very active ionosphere, in particular for the first $\sim1$~hr of the observation where the TEC values change very rapidly (more typical values are $\lesssim 0.2$~TEC~units).}
\label{fig:clockandtec}
\end{figure*}

\begin{figure*}
\begin{center}
\includegraphics[angle =0, trim =0cm 0cm 0cm 0cm,width=0.49\textwidth]{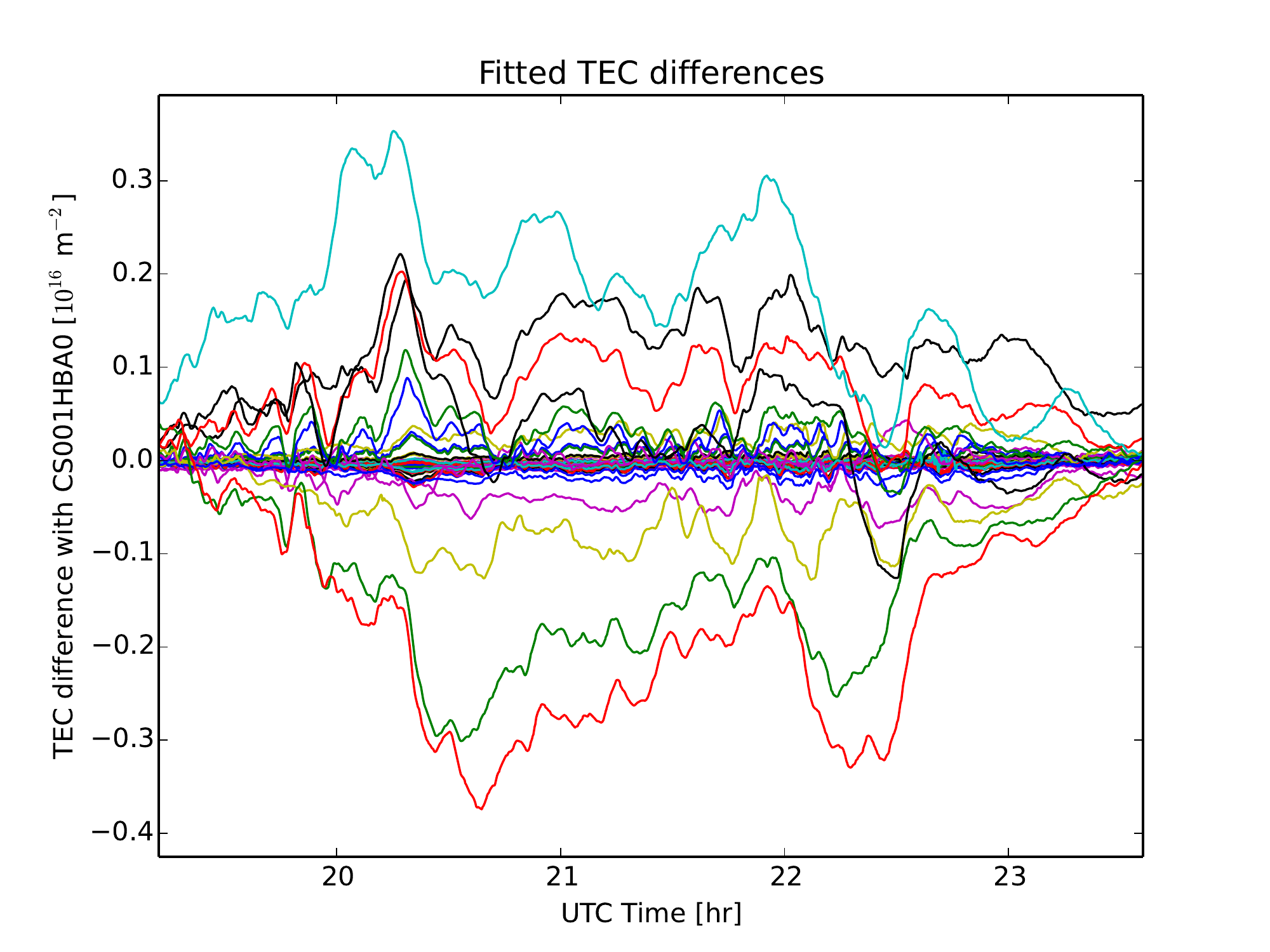}
\includegraphics[angle =0, trim =0cm 0cm 0cm 0cm,width=0.49\textwidth]{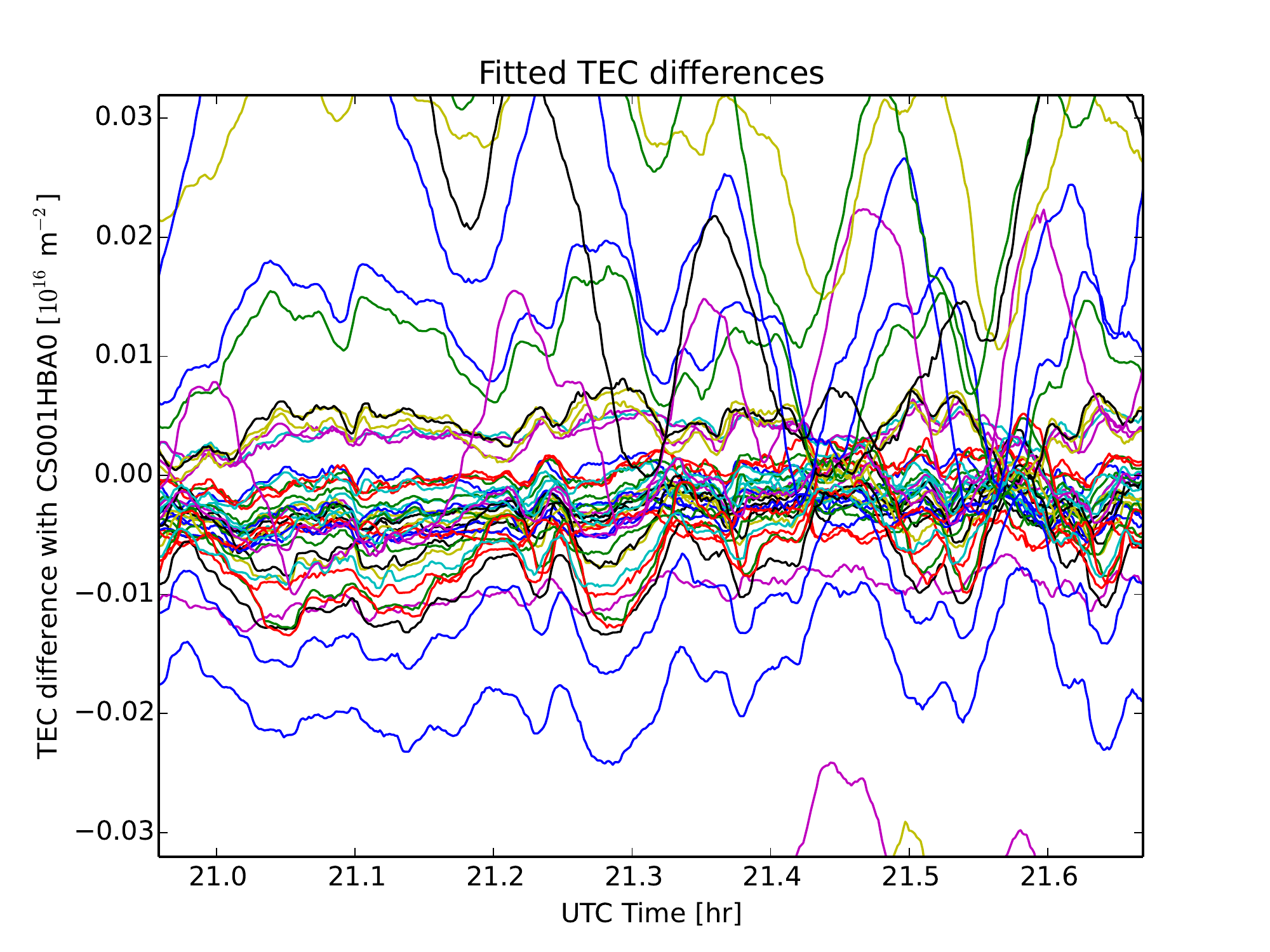}
\end{center}
\caption{TEC values with respect to reference station CS001HBA0. These two figures provide zooms for Figure~\ref{fig:clockandtec} (right panel). The ``band'' of lines with TEC values close to zero represent core stations that see more or less the same ionosphere as the reference station. The large TEC values and rapid changes require corrections on short (10~s) timescales to prevent ionospheric ``blurring'' of the image.}
\label{fig:teczoom}
\end{figure*}

The remote LOFAR stations  have their own clocks (to timestamp the data before correlation) which are not perfectly synchronized with the single clock that is used for all the core stations.  This causes a strong phase delay across the frequency band (phase $\propto \nu$) for the remote-remote and core-remote baselines. For our observation, very large clock offsets (delays) were present for some of the remote stations, up to 200~ns {(with respect to the core)}. In addition to these clock offsets, the clocks for the remote stations can drift  by $\pm 15$~ns over the course of an observation with respect to the core station clock. The large clock offsets need to be corrected (see Sect.~\ref{sec:DDE}), before we can proceed with the directional part of the calibration.

We developed a method to derive the clock values from an observation of a bright calibrator source. In our case this was 3C147 which has a large enough flux contribution on all baselines to obtain high signal-to-noise (S/N) gain solutions on 5~s timescales. We will refer to this method as ``Clock-TEC separation''.  Once the clock values are determined we can correct the target field data for this effect with {\tt BBS}.  Removing the time varying clock is equivalent to applying a frequency and time dependent phase correction to each station. In the next paragraphs we outline this method.

\begin{figure*}[t!]
\begin{center}
\includegraphics[angle =0, trim =0cm 0cm 0cm 0cm,width=0.49\textwidth]{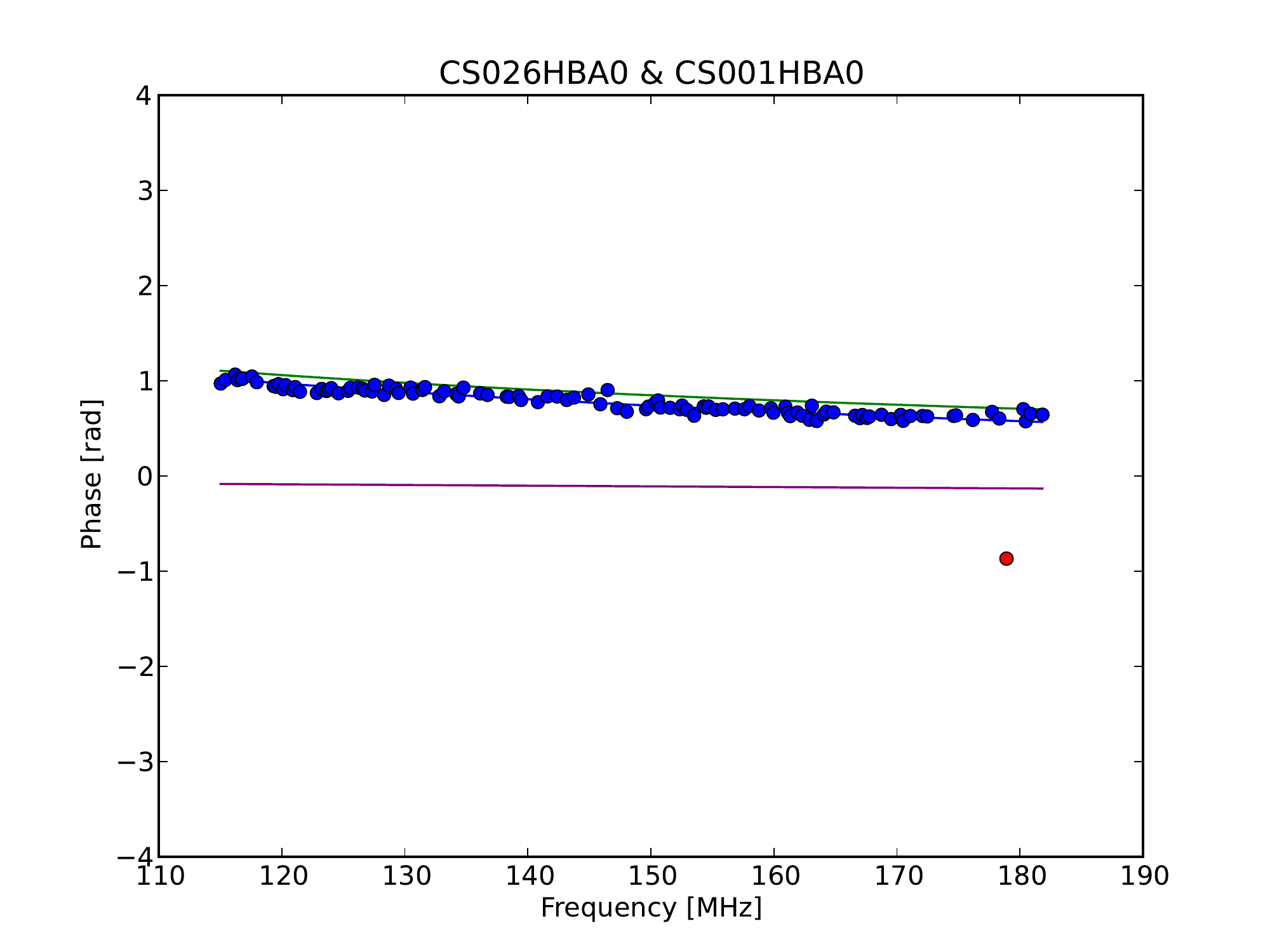}
\includegraphics[angle =0, trim =0cm 0cm 0cm 0cm,width=0.49\textwidth]{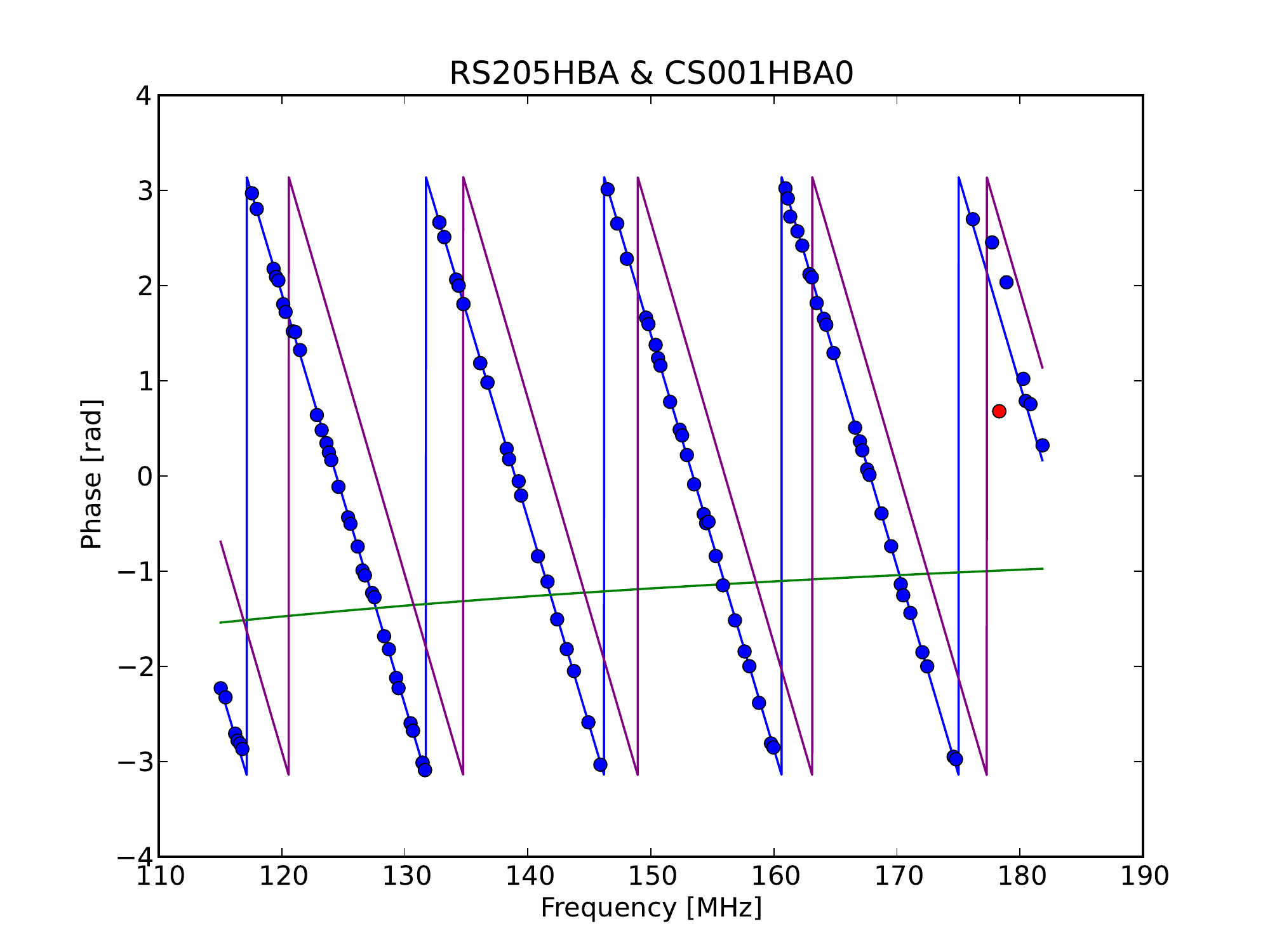}
\includegraphics[angle =0, trim =0cm 0cm 0cm 0cm,width=0.49\textwidth]{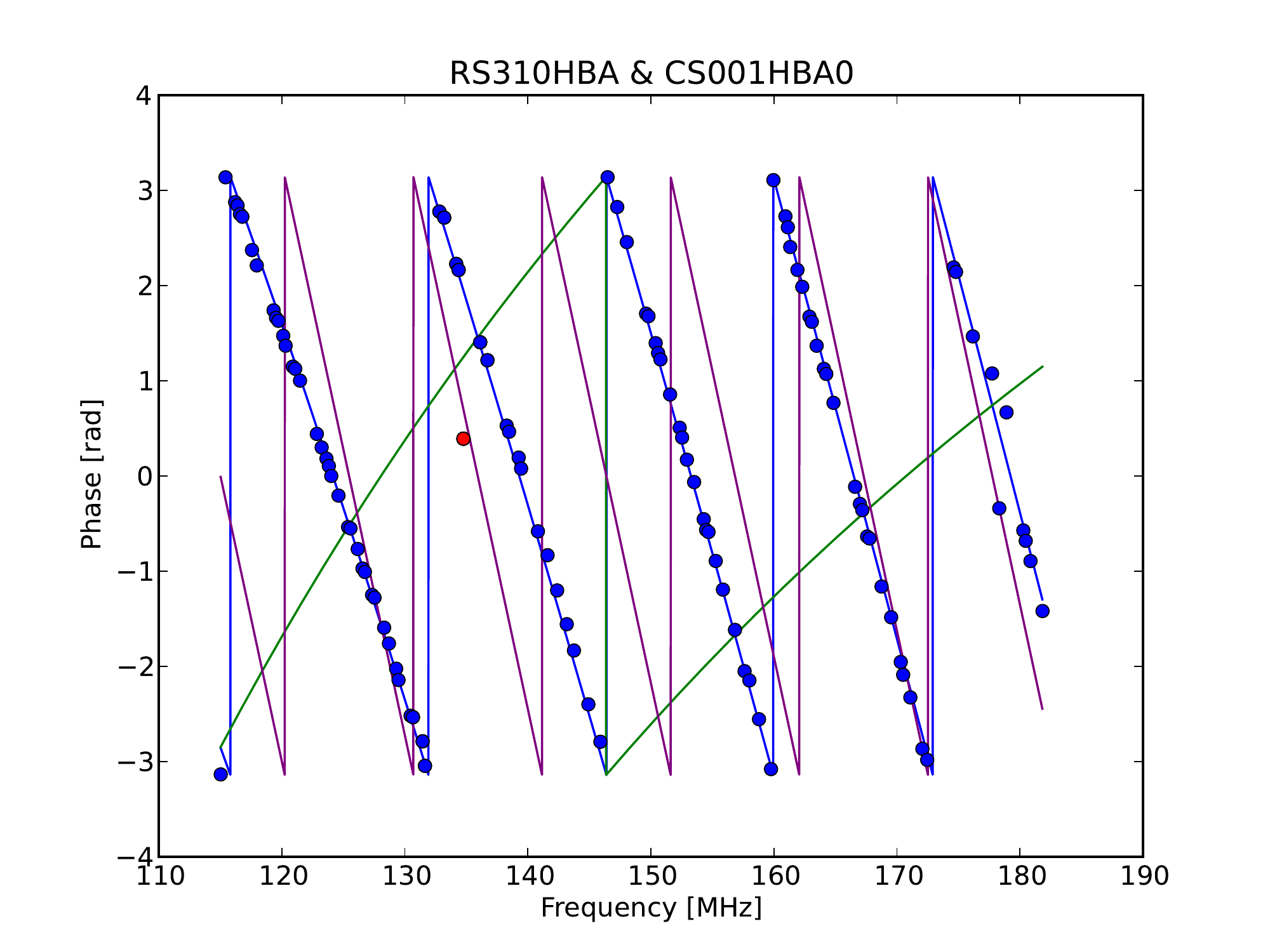}
\includegraphics[angle =0, trim =0cm 0cm 0cm 0cm,width=0.49\textwidth]{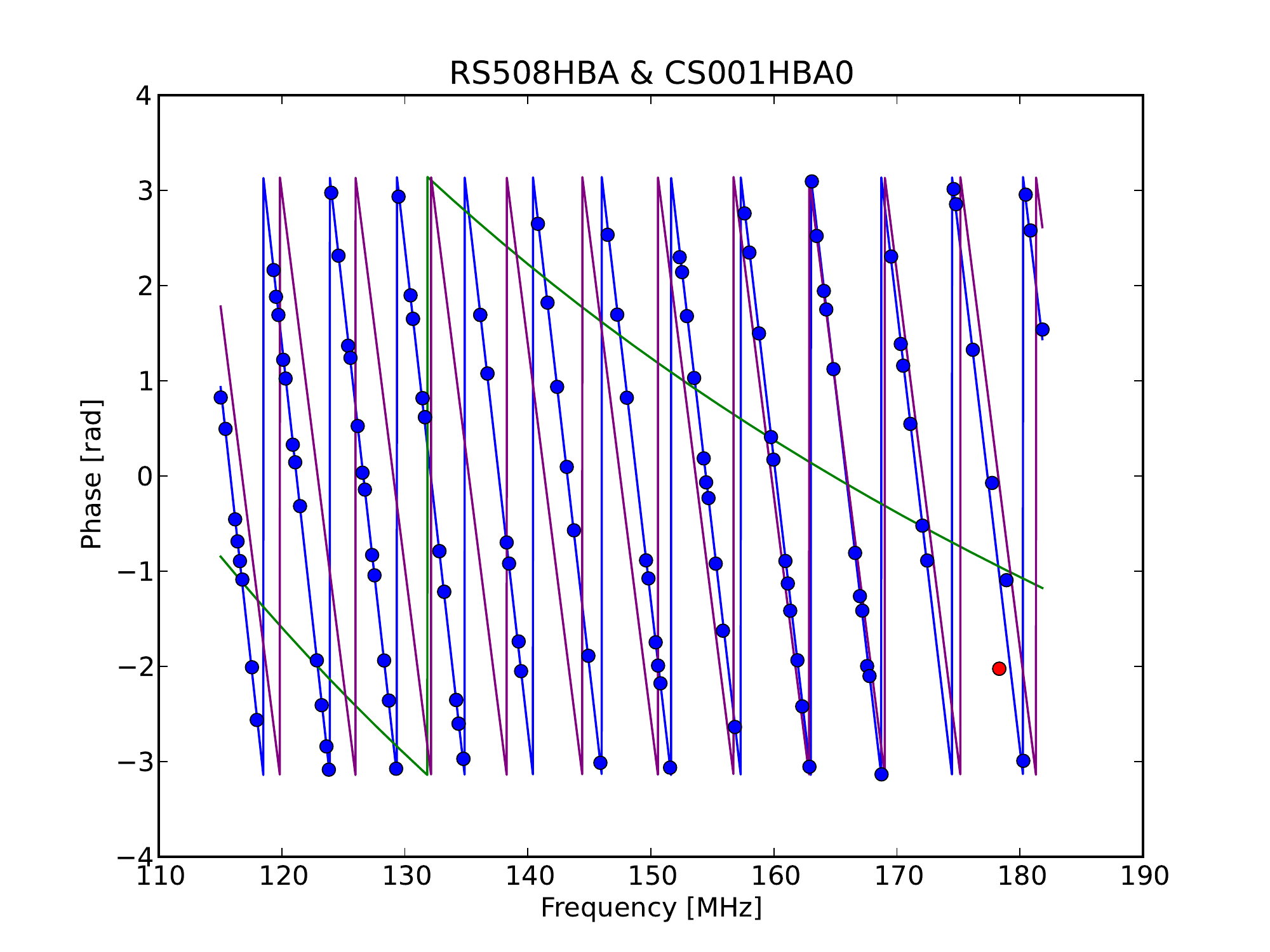}
\end{center}
\caption{Examples of the Clock-TEC fitting for one core and three remote stations. The distances to the reference station CS001HBA0  are $1.0$, $6.4$, $51.8$, $37.1$~km for CS026HBA0, RS205HBA, RS310HBA and RS508HBA, respectively. The blue points show the 3C147 phase  solutions as a function of frequency at an arbitrary 5~s time slot. Red points are outliers that were iteratively rejected during the fitting. These bad solutions are usually caused by subbands that are strongly affected by RFI. The blue line is the fitted model to the data. The separate contributions of the TEC and clock components to fitted phases are shown with green and purple lines, respectively. From these fits it can be seen that the very large clock offsets for the remote stations cause the majority of the phase changes across the HBA band. For the core station CS026HBA0 the fitted clock contribution is much smaller than for the remote stations. This is expected as all core stations operate in a single clock. The TEC component for CS026HBA0 is small as well, because the core stations are close enough together that they see more or less the same ionosphere.}
\label{fig:freqphase}
\end{figure*}

A complication for low-frequency radio observations is that the phases also vary because of the ionosphere and that these phase variations are not linear across the observed bandwidth. When we are observing far enough from the ionospheric plasma frequency (which is typically located around $\sim 10$~MHz) the effect has a frequency dependence of:  phase~$\propto \nu^{-1}$. The amount of phase change is directly related to the free electron column density along a line of sight through the ionosphere. The unit for TEC  is $10^{16}$~m$^{-2}$. For an interferometer (and neglecting polarization) only the differences in TEC (and clock) are relevant. The observed phase difference for a baseline can be thus written as

\begin{equation}
\Delta\rm{phase}(\nu, t) = 2 \pi p_0(t) \nu - \frac{8.448 \cdot 10^{9} p_1(t)} {\nu}  \mbox{  [rad]  ,}  
\label{eq:io}
\end{equation} 
where $p_0$ is the clock difference and $p_1$ the  TEC difference (the constant of $8.448 \cdot 10^{9}$ relates TEC to phase). In principle we could directly solve for the clock and TEC differences on the visibility data, employing the fact that the phase frequency dependence of clock and ionosphere are different. This requires that there is sufficient frequency coverage and/or signal-to-noise to detect the non-linear behavior (the $p_1/\nu$ term) of the ionosphere in Eq.~\ref{eq:io}. We decided not to directly solve Eq.~\ref{eq:io} on the visibility data, because this would require including all 121 subbands for the solve. This is computationally  expensive and that also made it less practical for testing the clock-TEC separation approach.

Instead of directly solving for clock and TEC on the visibility data, we use the 3C147 phase solutions. We thus have 121  phase solutions along the frequency axis as input to fit  clock and TEC difference via Eq.~\ref{eq:io} for every 5~s time slot. 
We do this fitting in {\tt python} using {\tt scipy.optimize.leastsq}. A difficulty with the this  least-squares solve is that Eq.~\ref{eq:io} has many local minima in $\chi^2$ space and the actual global minimum is not much deeper than neighboring local minima. In other words, our frequency coverage is not wide enough to easily separate the non-linear behavior of the TEC from the clock, unless good initial guesses are available. 

We use a brute force search on a grid to find clock and TEC initial guesses that are close enough to the actual solutions to achieve convergence. 
As expected, for the core stations the clock differences were close to zero (less than 1~ns for most stations), see Figure~\ref{fig:clockandtec}. The fitting was done separately for the XX and YY phase. For a few stations, we found a small but constant offset between the XX and YY phases. We determined these offsets by taking the median phase difference between the XX and YY phases over the entire length of the observation for each station\footnote{{We do not know the origin of these XX--YY phase offsets.}}. After the offsets were taken out, the resulting clock and TEC values were averaged and smoothed with a running median filter (with a window size of 15~s), see Figures~\ref{fig:clockandtec} and \ref{fig:teczoom}.

Based on the TEC solutions, we decided to flag the first 1.2~hrs of the data since the  TEC differences were very large and changed rapidly over time. During this time period the phase rate exceeded more than 1~rad per 10~s on the longest baselines. To avoid time decorrelation we  discarded the first 1.2~hrs of data for the rest of the processing.

\subsection{Clock and amplitude transfer}
The clock values and XX--YY phase offsets were transferred, together with the amplitudes, to the target field data. The amplitudes were also inspected for outliers and smoothed along the time and frequency axis with a running median filter (with a window size of 5 min along the time axis and 3 subbands along the frequency axis). After applying these corrections, the resulting target field data was free of clock delays and the visibility amplitudes were in units of jansky. We did not transfer the TEC values because these are direction dependent and differ for the calibrator and target fields. 

For the LOFAR observations discussed in this work, simultaneous calibrator observations were available for the entire length of the observing run. However, this is not required and a short $\sim 10$~min observation of a primary calibrator source at the start or end of the main observing run is also sufficient to transfer the clock and amplitudes. The reason for this is that the amplitudes are usually stable over the length of an observation, and the clock drifts remain within $\pm \sim15$~ns of the global offsets. The remaining clock drifts are small enough to be corrected for at a later stage (during the self-calibration or the direction dependent calibration).

\subsection{Self-calibration of the target field data}
\label{sec:selfcal}
We combined the clock and amplitude corrected individual subbands of the target field into groups of 10. The resulting combined datasets thus have about 2~MHz bandwidth and 20 channels. It is assumed that the phase change across 2~MHz bandwidth is small enough to be neglected during the calibration. As can be judged  from Figure~\ref{fig:freqphase} (green lines) this a reasonable assumption since the phase change due to the TEC is much smaller than 1~rad across a 2~MHz band.  The concatenation of 10~subbands also increases the S/N for the calibration by a factor of about $\sqrt{10}$.
We performed another round of RFI removal with {\tt AOFlagger} on the combined 10~subband datasets as this allowed the detection of lower-level RFI. {Typically, 95\% of the target field data remained after this second round of RFI excision}. We discarded the data between 112 and 120~MHz, as signal to noise was relatively poor there compared to the data above 120~MHz. 

We phase-calibrated these 10~subband datasets against the GMRT 150~MHz model described in Sect.~\ref{sec:ateam} which has a resolution of $26\arcsec\times22\arcsec$ {and a r.m.s. noise that varies mostly between 1 and 2~mJy~beam$^{-1}$.} The phase calibration was carried out on a 10~s timescale to avoid phase decorrelation by the ionosphere.  A spectral index of $-0.8$ was assumed to scale the GMRT model to the different frequencies. The S/N in the 10~subband datasets was high enough to obtain good quality phase solutions for all stations, i.e., we could easily track the phase solutions over time. The station beam model was used when computing the sky model. We then imaged these 10~subband datasets with {\tt CASA} and corrected for the effect of the station beams in the phase center. This does not allow for a full proper beam correction across the  FoV when imaging but is enough to allow for subsequent self-calibration cycles.  

From these  {\tt CASA} images we created a new sky model with {\tt PyBDSM}. This model is an apparent sky model as it is not fully corrected for the station beams (except in the phase center). We carried out an amplitude and phase (self)calibration against this model and again re-imaged the data. All the imaging was done with outer uv-range cut of {7~k$\lambda$}, limiting the resolution to $\sim25\arcsec$, a FoV of $\approx13\degr$, and \cite{briggs_phd} weighting ({\tt robust=0}).  The resolution and FoV imaged are a compromise between the accuracy of the sky model obtained and processing speed.  {For the solving step we included all available baselines.} W-projection \citep{2008ISTSP...2..647C,2005ASPC..347...86C} was employed to deal with the non-coplanar nature of the array. Furthermore, during the imaging clean masks were used, with the mask derived from a previous imaging run without mask. The clean masks were made with {\tt PyBDSM}, detecting islands of emission with a $3\sigma_{\rm{rms}}$ island threshold and a pixel threshold of $5\sigma_{\rm{rms}}$. A box size of $70\times70$ pixels was used to compute the locally varying r.m.s. across the maps to take into account (calibration) artifacts around strong sources.

The products of the above steps are apparent-flux ``medium resolution'' images of the sky between 120 and 181~MHz in steps of about 2~MHz (Figure~\ref{fig:directions}). From this process we obtained a total of 29 of these images. Two blocks of 10~subbands were excluded because they were affected by strong RFI, see Table~\ref{tab:observations}.

{
\subsubsection{Accuracy of the flux-scale bootstrapping}
\label{sec:3C147bootstrap}

To assess the accuracy of the transfer of the flux-scale from the calibrator 3C147 to the target field, we extracted catalogs from the GMRT and medium resolution LOFAR image at 150~MHz with {\tt PyBDSM}. In this case, the HBA image was produced with {\tt awimager} \citep{2013A&A...553A.105T} which fully corrects for the time-varying primary beam. We used the same box size of $70\times70$ pixels to compute the r.m.s. map. We cross-matched the catalogs using a matching radius of 5\arcsec. Sources with a S/N less than 10 or that were significantly resolved (a ratio of integrated flux over peak  flux larger than 2) were excluded.

The ratio of the HBA over GMRT integrated flux densities is shown in Figure~\ref{fig:mediumresfluxcomp}. The ratio seems to be constant as a function of radial distance to the pointing center. From the medium flux ratio we determine a difference in flux scale of about a factor 0.8 between LOFAR and the GMRT. Such differences are not unexpected given the uncertainties in the calibration of the low frequency flux-scale with the GMRT \citep[which is estimated to be about 10\%,][]{2011A&A...535A..38I} and the complex LOFAR station beams \citep{2012A&A...543A..43V,2014ApJ...793...82V}. Recently, it has been found that the normalization of the LOFAR HBA beam was not correctly implemented in the models (work is ongoing to fix this issue). {When transferring a flux-scale from one pointing to another, differences of the order of what we find are expected \citep[based on work carried out as part of the  MSSS survey,][]{2015A&A...582A.123H}. We therefore suspect that most of the difference with the GMRT flux-scale is related to this LOFAR beam model issue. However, at the moment it is not yet possible to calculate what the expected error is for our specific observation. We take a conservative approach here and} assume that the GMRT flux-scale is correct and use a scaling factor of 1.2 to correct our extracted flux densities from the LOFAR HBA maps.

\begin{figure}[th!]
\begin{center}
\includegraphics[angle =180, trim=0cm 0cm 0cm 0cm,width=0.45\textwidth]{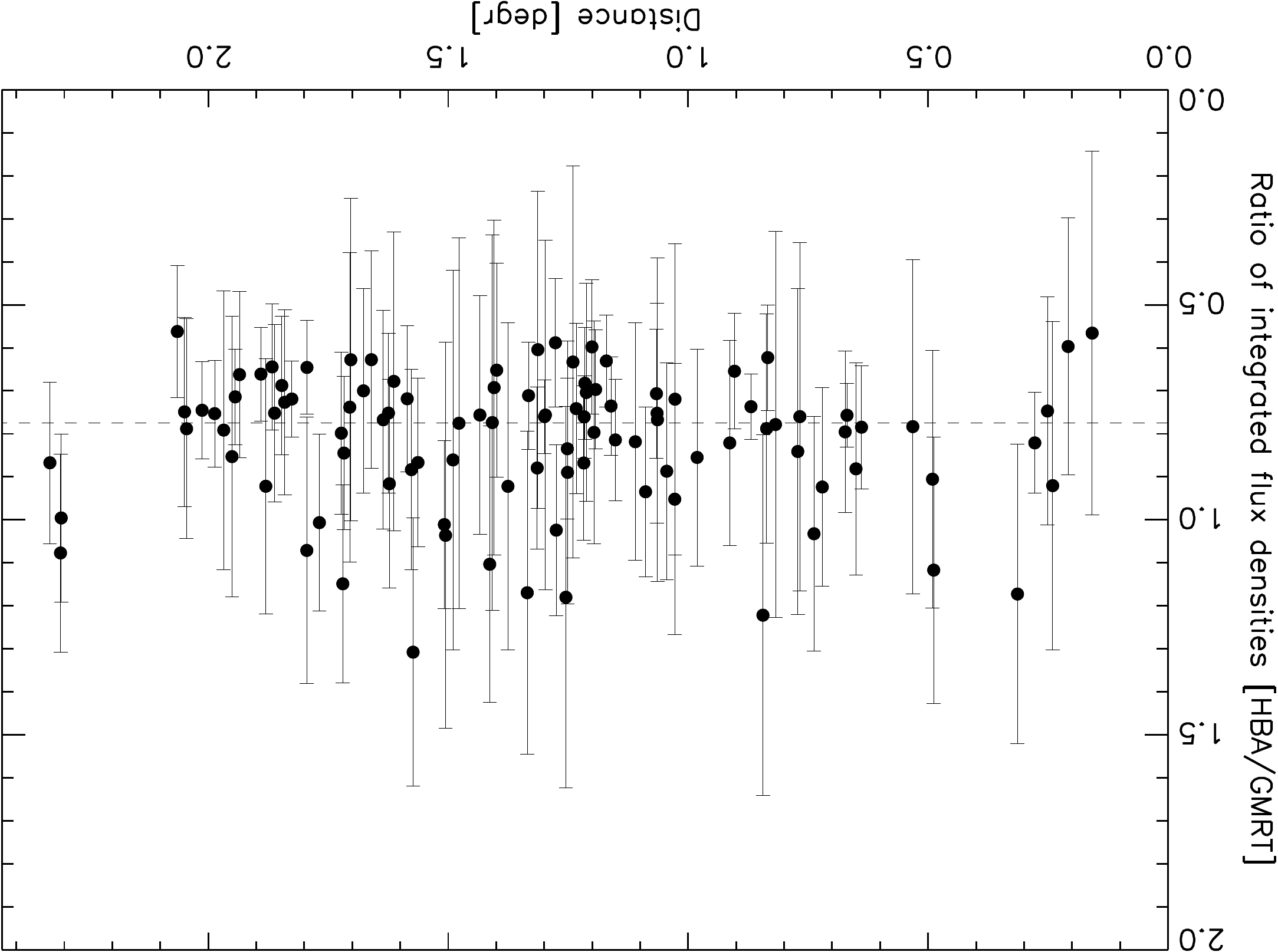}
\end{center}
\caption{{Integrated flux density ratios (LOFAR HBA over GMRT) for sources in the Toothbrush field at 150~MHz as a function of distance to the pointing center. The dashed line indicates the median flux density ratio.}}
\label{fig:mediumresfluxcomp}
\end{figure}
}

\subsection{Subtracting the sources from the data}
\label{sec:selfcalsub}

As a final step, we subtracted the sources from the 10~subband datasets. This was done using a two step approach. First we subtracted the clean components from the medium resolution images (with the corresponding gain solutions). We then re-imaged this clean component subtracted data at a lower {$\sim1.5\arcmin$~resolution using an outer uv-range cut of 2~k$\lambda$}, but now with a FoV of $\approx33\degr$. {This allowed us to detect sources in the first and second sidelobes as well as extended low-surface brightness emission in the field that was not cleaned in the medium resolution imaging.} The clean components found in the low-resolution image were again subtracted {with the direction independent self-calibration solutions. The reason for imaging this far out is to (i) remove the contribution of sources in the far sidelobes, so they do not influence the direction dependent calibration and (ii) search for the presence of ``off-axis'' sources  that are bright enough that they need to be included in the direction dependent calibration (often 3C sources).} 

Completing the above steps, we are left with 29 2-MHz-wide datasets with all sources subtracted and corresponding sky models. The 29~sky models take care of the frequency dependence of the sky at this point. Note that these datasets do not have the gain solutions applied, rather the clean components were   ``corrupted'' with the gain solutions and subtracted from the uncorrected data. These ``empty'' datasets serve as input for the direction dependent calibration scheme.

\section{Correction for direction dependent effects:  ``facet calibration''}
\label{sec:DDE}
After self-calibration, significant artifacts remain around (bright) sources in LOFAR images, see Figure~\ref{fig:directions}. Also the r.m.s. noise levels in the 10~subband medium resolution images are a few mJy~beam$^{-1}$. This is a factor of 5--10 higher than the expected thermal noise with these imaging settings. The increased noise and artifacts are caused by DDEs, namely the station beam and the ionosphere. 

Previous work has shown that gain corrections towards $\sim 10^2$ directions are needed to correct for these errors  \citep{2013A&A...550A.136Y}. A concern here is that this direction dependent calibration involves solving for many parameters. Ideally one should try to keep the number of degrees of freedom (d.o.f.) in the calibration small with respect to the number of independent measured visibilities. There is also the requirement of having enough S/N in each direction to enable an accurate calibration. The accuracy and completeness of the sky model to calibrate against, is also important. This is particularly relevant for LOFAR, because no high-resolution models of the sky are available at these frequencies. For a more detailed discussion on these various topics see \cite{2009A&A...501.1185I,2011A&A...527A.108S,2013MNRAS.435..597K,2013MNRAS.430.1457K,2015MNRAS.449.4506Y}. To keep the number of d.o.f. small during the calibration and achieve sufficient S/N, we make the following (reasonable) assumptions:

\begin{figure}[t!]
\begin{center}
\includegraphics[angle =0, trim =0cm 0cm 0cm 0cm,width=0.49\textwidth]{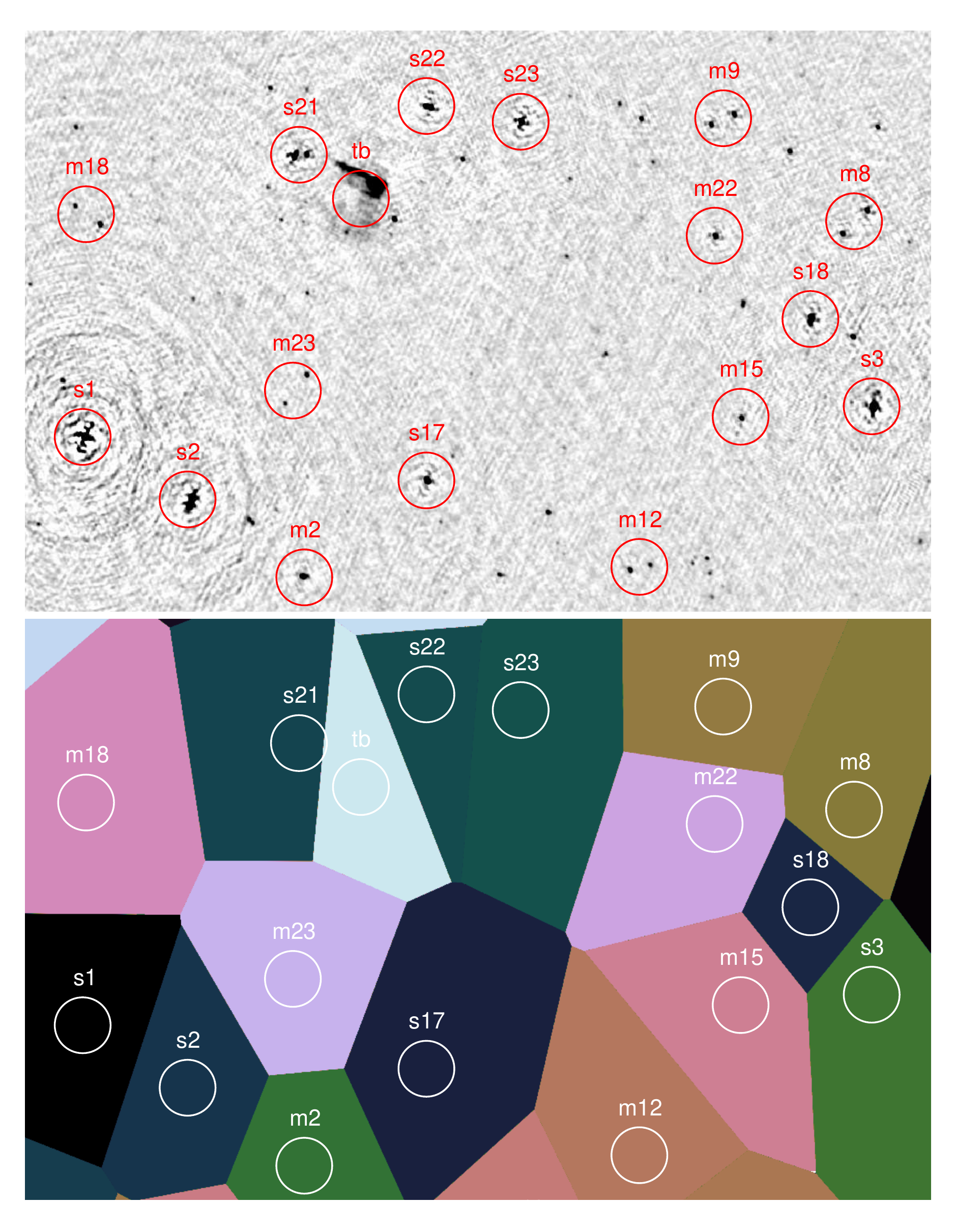}
\end{center}
\caption{Example of the calibration directions on top of a 150--152~MHz image (26\arcsec~resolution) displaying a region around the Toothbrush cluster (top panel).
Note that calibration artifacts are visible around the brighter sources as only self-calibration has been performed and no DDE calibration has been applied, see Sect.~\ref{sec:selfcal}. The FoV shown  measures $2.1\degr\times1.4\degr$. The center of each circle defines a direction towards a bright radio source (group). {Direction m9 is an example of a ``group''.} Based on these directions the sky is partitioned via a Voronoi tessellation scheme (bottom panel). The different colors associated with the regions are arbitrary and just for visual representation.}
\label{fig:directions}
\end{figure}

\begin{itemize}
\item the station beams vary slowly over time and frequency \citep{2013A&A...550A.136Y};
\item differential Faraday rotation can be neglected in the HBA band (unless extremely high dynamic range is required), meaning that the ionosphere affects the XX and YY phases in the same way;
\item the frequency dependence of the phase is: phase $\propto \nu^{-1}$  (note that the effects of the clocks have been taken out);
\item no other calibration errors are present besides ionosphere and beam;
\item all the direction dependent effects vary {smoothly across the FoV \citep[e.g.,][]{2009AJ....138..439C}};
\end{itemize}

{The last assumption is particularly relevant as it implies that we can divide up the sky into a number of  ``isoplanatic patches'' \citep[e.g.,][]{1984AJ.....89.1076S}.}


In the next subsections we outline the direction dependent calibration scheme, which we will refer to as ``facet calibration''. The scheme has some similarities to {\tt SPAM}  \citep{2009A&A...501.1185I,2014ascl.soft08006I} and {\tt Sagecal} \citep{2008arXiv0810.5751Y,2011MNRAS.414.1656K}, although there are also a few important differences (see Section~\ref{sec:comparisonspamandsage}). {Another technique that was developed to correct for the ionospheric phase errors is ``field-based calibration'' \citep{2004SPIE.5489..180C, 2005ASPC..345..337C}. For field-based calibration, snapshot images of bright sources are made and their position offsets are measured. With these measurements, an ionospheric model is fit which is subsequently applied during the imaging to correct for the source movements. However, field-based calibration methods are not suitable for arrays with very long baselines such as LOFAR \citep[][]{2005ASPC..345..399L,2009A&A...501.1185I}.}

\begin{figure*}[th!]
\begin{center}
\includegraphics[angle =0, trim=0cm 0cm 0cm 0cm,width=0.65\textwidth]{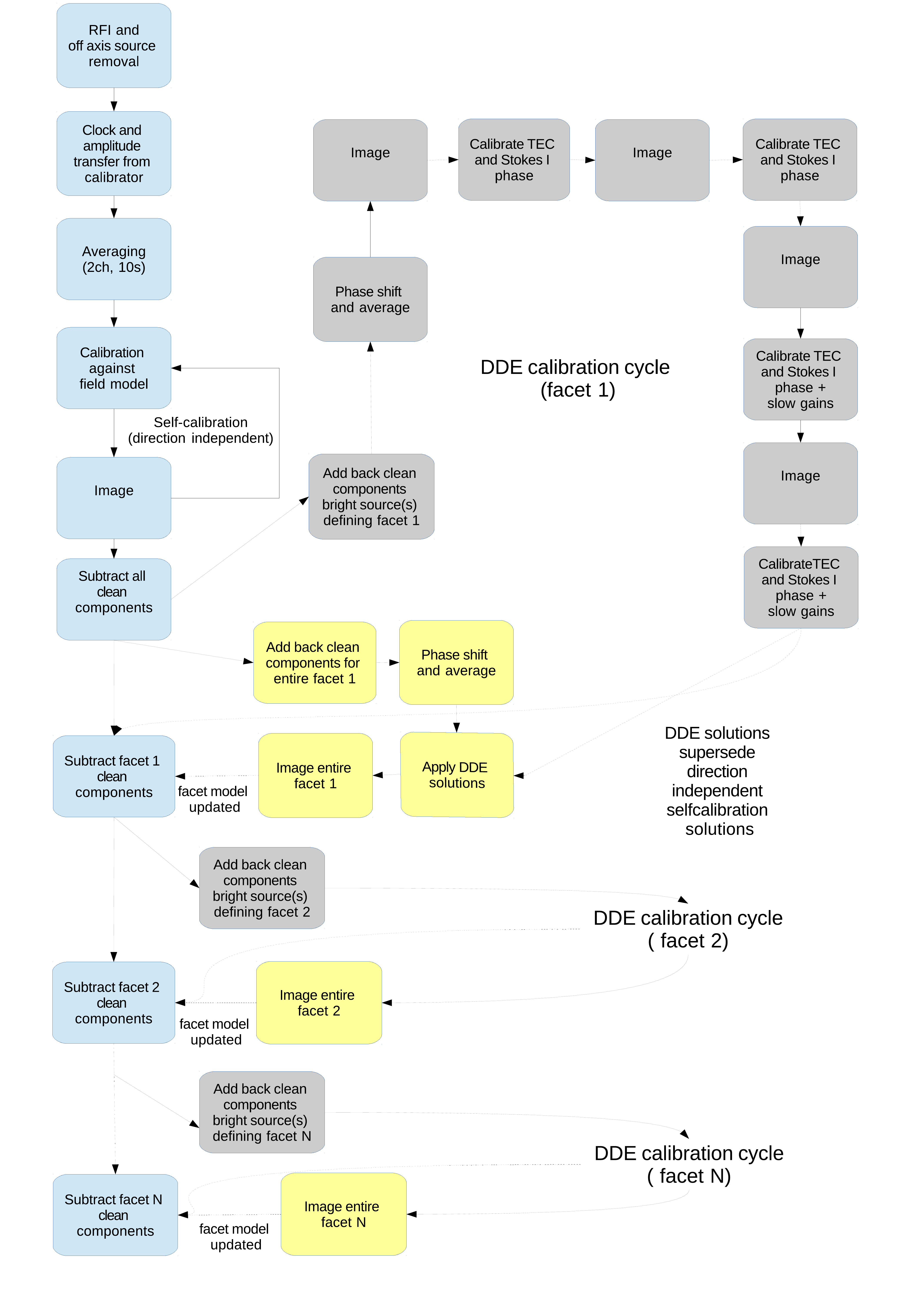}
\end{center}
\caption{Schematic overview of the HBA calibration scheme employed for this work. Gray colored boxes depict the DDE calibration cycle and yellow boxes the imaging of the facets.}
\label{fig:scheme}
\end{figure*}

Below we discuss this facet calibration scheme in more detail, in particular focussing on the parameters that are solved for during the calibration. A schematic overview of the calibration scheme is given in Figure~\ref{fig:scheme}.

\subsection{Dividing up the sky in facets}
A first step in the directional calibration is to divide up the sky into facets. When the sky is divided up into facets we make the assumption that the DDE calibration solutions towards the bright source (group) apply to the facet as a whole. {The ``center points'' of the facets are located on bright sources, or the approximate center of a group of closely separated (less than a few arcmin) bright sources. The number of facets required depends on (i) the specific field, (ii) ionospheric conditions and station beam shapes, (iii) the required dynamic range or noise level, and (iv) the science aim.  The considerations for the choice of calibrator directions, which define the facet layout, is very similar to, for example, {\tt Sagecal} or {\tt SPAM}. The main consideration is having sufficient flux available for calibration and the complexity of the sources (for example, very extended sources might require multi-scale clean which slows down the deconvolution steps).} 

{The selection of the center points of the facets is done by the user (but see Section~\ref{sec:future}).  An apparent flux density of at least $\sim 0.4$~Jy is required to define a center point. This is determined by the need to obtain direction dependent solutions with sufficient S/N. Center points are selected by visually inspecting the  25\arcsec~resolution images from the direction independent self-calibration. Naturally, the sources which show the strongest calibration artifacts (typically the brightest sources) end up in the user defined list of center points. In the case of a source group, the approximate center position of such a group is taken.}  

{For the Toothbrush field this resulted in a list of {67} center points (i.e, directions), which cover an area of about $1.5-2 \times$~the HPBW of the station beam. 
Fewer directions are defined beyond the HPBW because the number of sources with an apparent flux density of $>0.4$~Jy decreases steeply beyond this radius. Two  bright outlier sources (3C147 and 3C153),  located at radial distances $>8\degr$ from the pointing center, were also included for the Toothbrush field. This number directions is of the same order as used by \cite{2013A&A...550A.136Y}.  We also included an additional 20 directions with $< 0.4$~Jy of flux density beyond the HPBW\footnote{These were later discarded and were used to determine the limiting flux density (of $\sim 0.4$~Jy) for calibration.}.
}

\begin{figure}[t!]
\begin{center}
\includegraphics[angle =0, trim =0cm 0cm 0cm 0cm,width=0.49\textwidth]{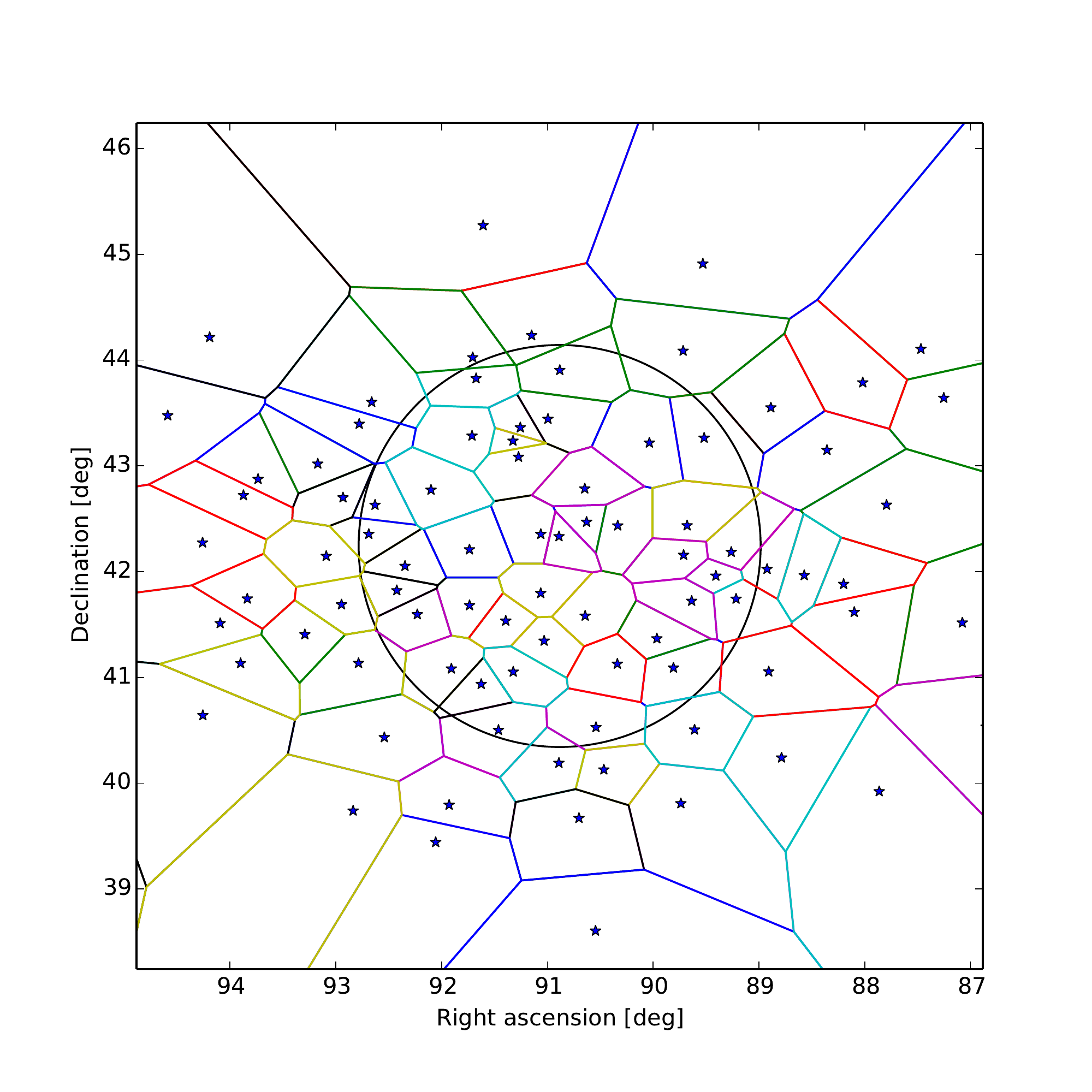}
\end{center}
\vspace{0mm}
\caption{{Voronoi tessellation used for the Toothbrush cluster field. The black circle indicates  the HPBW of the station beam at 150~MHz. Stars indicate the center points, placed on bright sources (or source groups), that define the tessellation.}}
\label{fig:voronoi}
\end{figure}

We then employ a Voronoi tessellation \citep[e.g.,][]{2000stca.conf.....O} scheme to make the facets; an example of this is given in Figures~\ref{fig:directions} and~\ref{fig:voronoi}. {This tessellation scheme assigns each point on the sky to the closest calibrator source (group). The area covered by facets is limited by the maximum image size allowed by the user for a given facet. This is done to prevent facets from growing too large. A consequence of this is that the user must take care to have a reasonably uniform distribution of calibrator directions in order to avoid the appearance of gaps in the final image.}

\subsection{Adding back the bright source (group)}

\begin{figure}
\begin{center}
\includegraphics[angle =0, trim =0cm 0cm 0cm 0cm,width=0.43\textwidth]{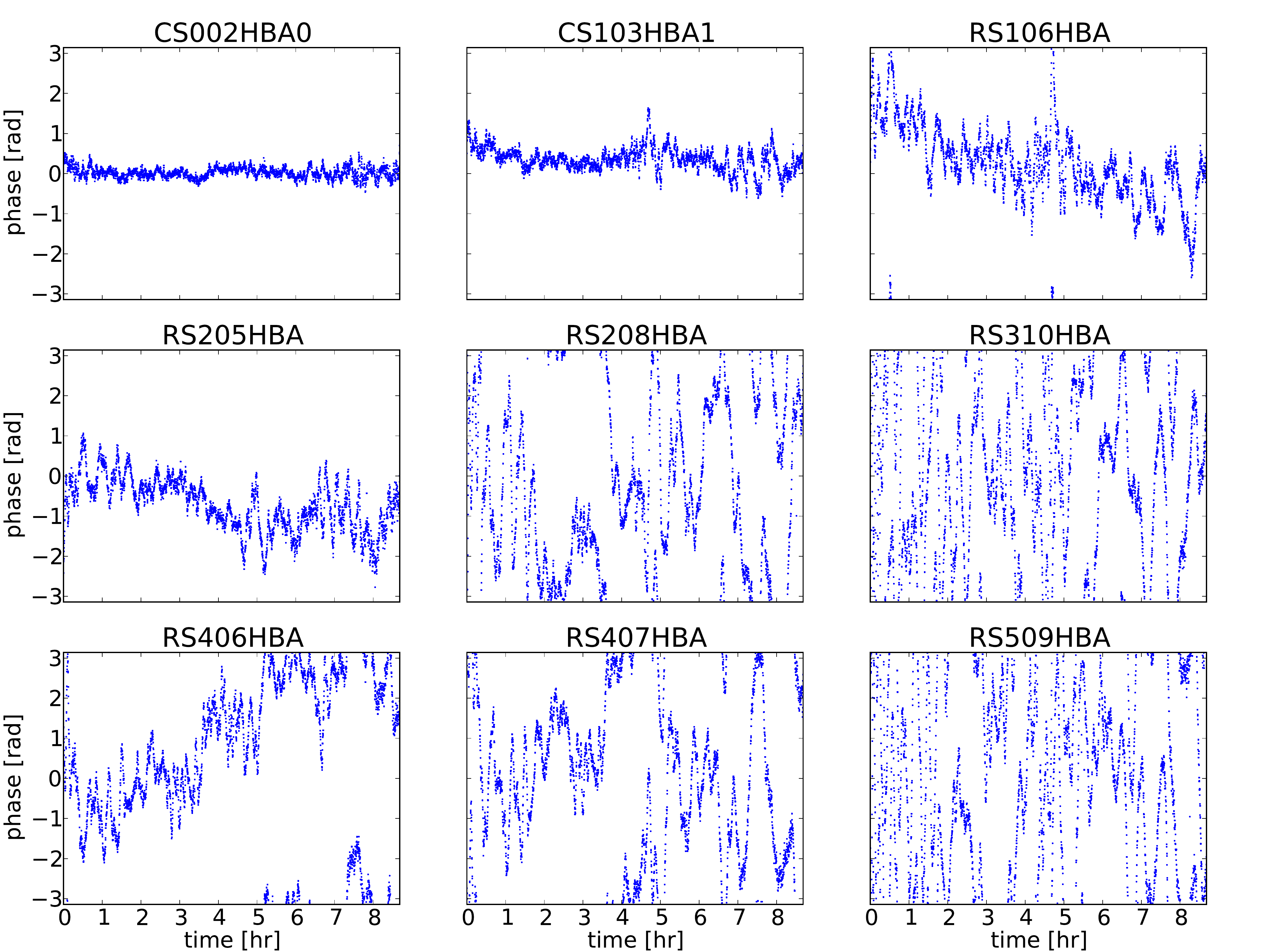}
\includegraphics[angle =0, trim =0cm 0cm 0cm 0cm,width=0.43\textwidth]{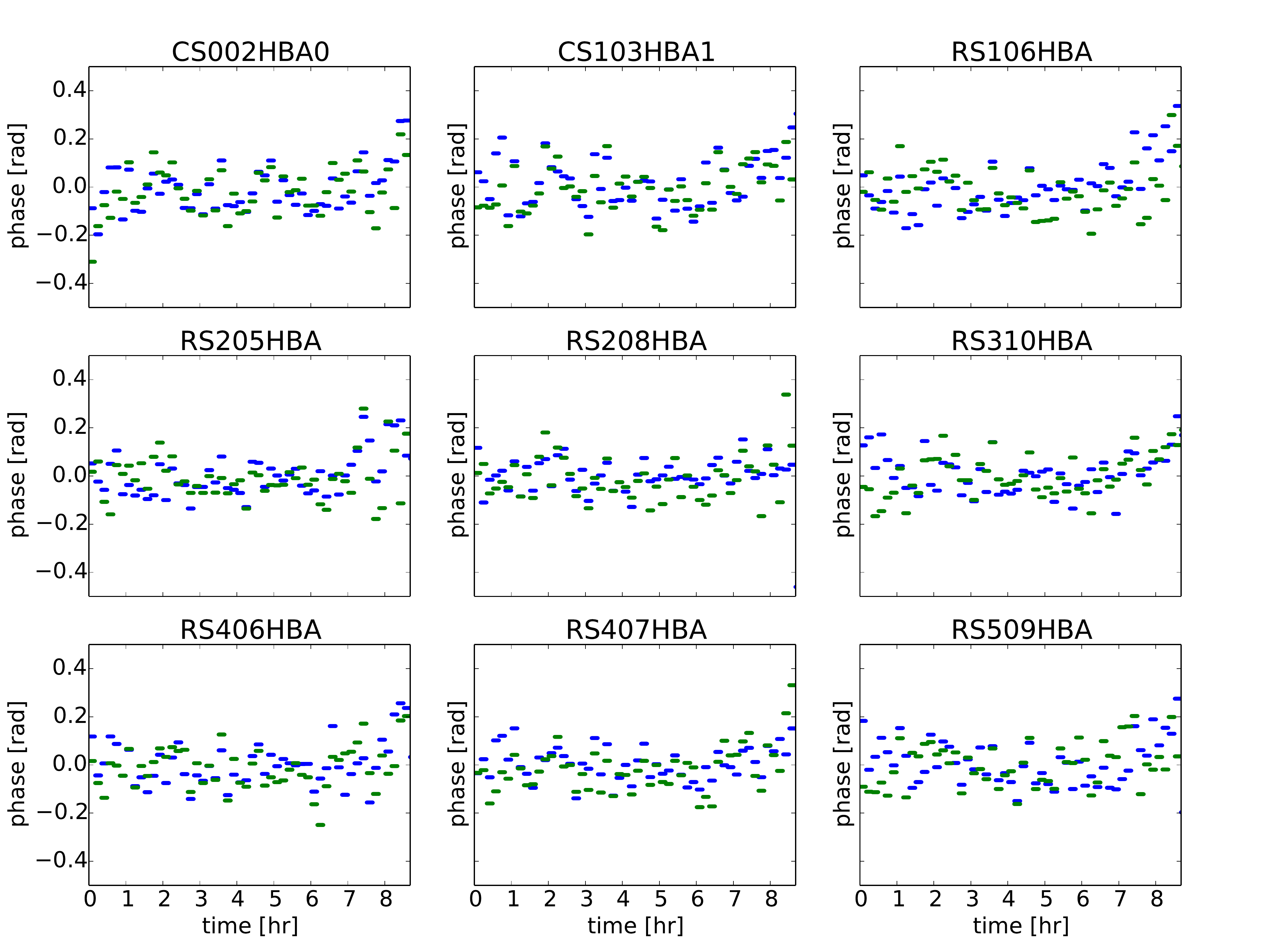}
\includegraphics[angle =0, trim =0cm 0cm 0cm 0cm,width=0.43\textwidth]{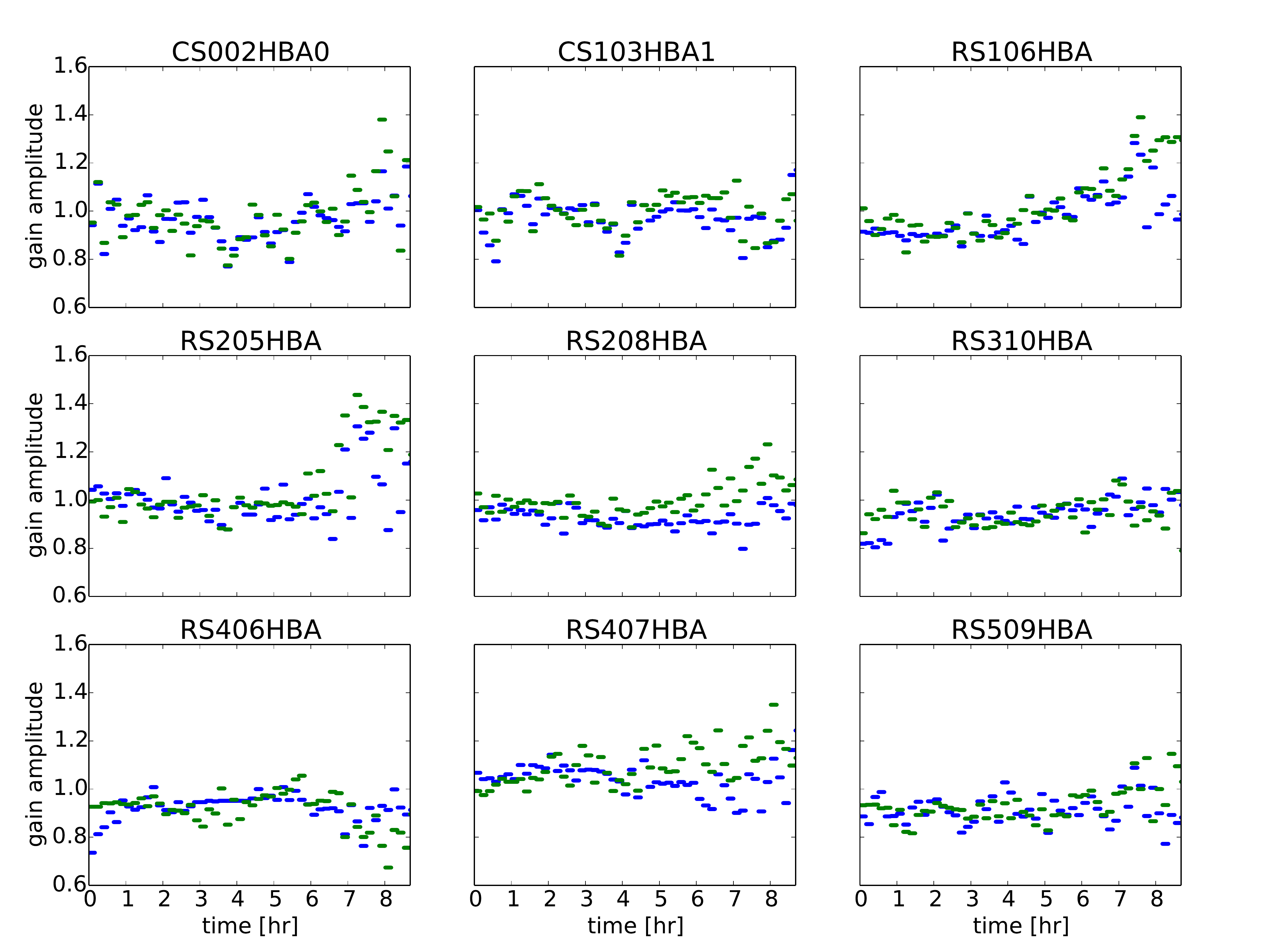}
\vspace{-3mm}
\end{center}
\caption{DDE solutions towards direction s21, see Figure~\ref{fig:directions}. Top: The effective {Stokes I} phase corrections are shown at a frequency of 150~MHz. For the images corresponding to this direction see Figure~\ref{fig:DDEimages}. Because we fit for TEC (and a phase) the actual corrections can be evaluated at an arbitrary frequency. The solutions are obtained on a timescale of 10~s using  20~MHz of bandwidth. Middle: XX (blue) and YY (green) phase solutions for the 150--152 MHz subband block.  The solutions are obtained on a timescale of 10~min. Phases are always plotted with respect to core station CS001HBA0. Bottom:  XX (blue) and YY (green) amplitude solutions corresponding to the middle panel. The distances to CS001HBA0 are 0.37, 2.0, 8.8, 6.4, 27.1, 51.8, 14.3, 20.8, and 55.7~km for CS002HBA0, CS103HBA1, RS106HBA, RS205HBA, RS208HBA, RS310HBA, RS406HBA, RS407HBA, and RS509HBA, respectively.\vspace{0mm}}
\label{fig:s21sols}
\end{figure}

\label{sec:addbackDDE}
The next step in the scheme consists of adding back a bright source or source group (which defines the facet position) to the visibility data. Typically, the source covers an area of a few sq. arcmin, which is much smaller than the size of the facet created via the Voronoi tessellation\footnote{The main reason for not adding all the sources back in a facet is to speed up the (self)calibration cycle that obtains the DDE solutions in this direction.}. The data are then phase rotated to the position of the source that was added back and averaged down to a channel resolution of $\sim2$~MHz, so each 10 subband block is averaged down to 1 channel. The fact that the source (group) only covers a small area means that after phase rotation we can average much more in frequency without being affected by bandwidth smearing. This averaging step is crucial, as the size of the data is reduced by a factor of 20 which speeds up the subsequent calibration cycles (see the next Section). No time averaging is done because we need to correct for the ionospheric phase changes on short timescales. 

\subsection{Self-calibrating a bright source or bright source group}
\begin{figure*}[th!]
\begin{center}
\includegraphics[angle =0, trim =0cm 0cm 0cm 0cm,width=0.95\textwidth]{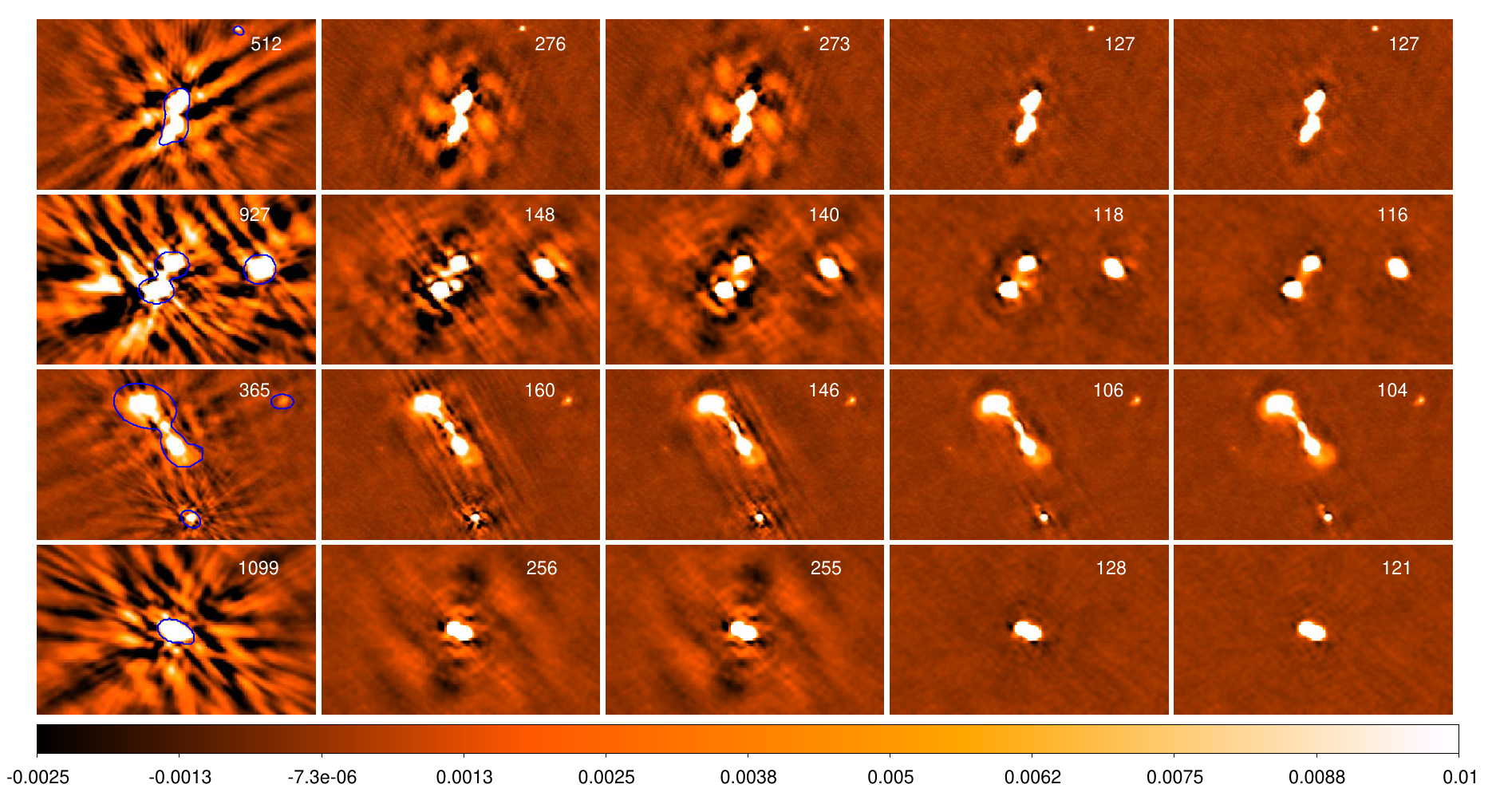}
\end{center}
\caption{Images showing the incremental improvements during the DDE calibration, see Sect.~\ref{sec:DDEselfcal}. For reference, the first and second row of images show direction s2 and s21, respectively (Figure~\ref{fig:directions}). All images are made using the full dataset (120--181~MHz, {\tt robust=-0.25}) and have a resolution of $8\arcsec\times6.5\arcsec$. Note that at this resolution many of the bright DDE calibrator sources are resolved. The first column displays the images made with the (direction independent) self-calibration solutions, see Sect~\ref{sec:selfcal}. {The blue contours show the clean mask that was created with {\tt PyBDSM} for the imaging. The clean mask is updated at each imaging step during the DDE calibration (not shown).} The next columns display improvements during the DDE calibration step (see also Figure~\ref{fig:s21sols}). Second column: first DDE TEC+phase iteration. Third column: second DDE TEC+phase iteration. Fourth column: third DDE TEC+phase iteration and first DDE XX and YY gain (amplitude and phase) iteration. Fifth column:  fourth DDE TEC+phase iteration and second DDE XX and YY gain (amplitude and phase) iteration. For all four directions the TEC+phases were solved for on a 10~s timescale. The XX and YY gains were solved for on a 10~min timescale, except for the source in the top row for which this was 5~min. The scale bar at the bottom is in units of Jy~beam$^{-1}$. The images in the first and third row were cleaned with multi-scale clean because of extended emission. The r.m.s. noise level in each of the images is indicated in the top right corner in units of $\mu$Jy~beam$^{-1}$.}
\label{fig:DDEimages}
\end{figure*}

\label{sec:DDEselfcal}
To obtain the DDE solutions a self-calibration cycle is performed on the bright source (group) that defines a particular facet. {This is similar to the procedure followed by {\tt SPAM}.}
At this time all the data (120--181~MHz) are imaged together using the MS-MFS clean algorithm \citep[][]{2011A&A...532A..71R} as implemented in {\tt CASA} with {\tt nterms=2}. Multi-scale\footnote{We use the {\tt CASA clean} option {\tt scales = [0,3,7,25,60,150]}.} clean \citep{2008ISTSP...2..793C} is employed for a few complex extended sources (Figure~\ref{fig:DDEimages}). Clean masking is done as described in Sect.~\ref{sec:selfcal}. {The imaging is carried out with all available baselines and no outer uv-range cut is imposed. To speed up the DDE calibration cycle and obtain a reasonable starting model we apply the direction independent self-calibration solutions (see Section.~\ref{sec:selfcal}) at the first imaging step. After we obtain the starting DDE calibration model the direction independent self-calibration solutions are discarded.}

The first two self-calibration cycles consist of Stokes I phase and TEC calibration on the visibility data using {\tt BBS}. At this point, instead of having 29~sky models for each of the 10~subband blocks, we now go down to a single clean component model that is valid over the entire 120--181~MHz band (each clean component having a flux and a spectral index, i.e., corresponding to {\tt nterms=2} in the {\tt CASA clean} task). Imaging the complete bandwidth at once has the additional advantage of an improved point spread function (PSF) as the large frequency coverage almost completely fills the uv-plane; in addition it allows cleaning fainter sources/emission that can then be included in the sky model. For most of the directions this phase and TEC calibration is carried out on the shortest timescale of 10~s to correct for the ionosphere. For about a {dozen} facets we increase this solution interval to 20~s because the available S/N is lower. 
For the calibration, a single TEC and phase parameter are found per station per 20 MHz bandwidth. We thus solve for only 6 parameters (3 phase and 3 TEC values, {see also Section~\ref{sec:dof}}) per antenna for the entire 120--181 MHz band. See Figure~\ref{fig:s21sols} (top panel) for an example of the calibration solutions.

After three rounds of this Stokes I phase + TEC self-calibration, we solve for gains on timescales ranging from 5 min to 20~min, based on the source (group) flux density. The phase + TEC corrections were pre-applied before solving for the XX and YY gains. This ``slow gain'' calibration is carried out to correct for the slowly varying station beams. The calibration is done independently per 10~subband block because the beam corrections are frequency dependent. The number of parameters solved for in this step is a factor of $\sim 20$ lower than what is typically done with {\tt Sagecal}, which solves for full Jones on a per subband basis on similar timescales. 

{We run a median window filter on the amplitude solutions to find and replace potential outliers. However, very discrepant amplitude solutions are rare at this stage in the calibration as no bad data should be present. Bad data should have been identified at earlier steps, i.e., using the calibrator observations or during the direction independent self-calibration.} 

The phase components of the ``slow gains'' are close to zero, i.e., $\ll 1$~rad (Figure~\ref{fig:s21sols}, middle panel). This is expected because the phase component of the beam variations are small within the main FoV and the ionospheric variations have already been taken out. The slow-gain calibration is followed by another round of Stokes-I phase and TEC calibration and another final round of XX and YY gain  calibration. The above scheme basically mimics a joint ``short-timescale phase+TEC'' and ``slow-timescale gain'' calibration. The reason for this approach is that {\tt BBS} cannot jointly solve for parameters on different timescales.

For the final solutions we normalize the global amplitudes to prevent the flux-scale from drifting. In all cases these normalization corrections were very small (a few percent or less). Thus when this self-calibration scheme has finished, we have obtained a set of solutions (Figure~\ref{fig:s21sols}) for a particular direction. The improvement in image quality, over the previous direction independent self-calibration (Sect.~\ref{sec:selfcal}) is very significant after completing this step. An example of the increase in image quality for the calibration scheme described in this section is shown in Figure~\ref{fig:DDEimages}.

{The images are not completely free from calibration artifacts: some small scale negative and positive artifacts (at levels of $\lesssim1\%$ of the peak flux) are visible very close to the bright calibrator sources. From tests we noticed that the magnitude of these residuals further decreases when adding additional calibration cycles following the scheme outlined above. However, due to computational limitations we decided not to increase the number of calibration cycles. These calibration artifacts could also partially have resulted from the imperfect subtraction of other sources, or indicate calibration errors on shorter timescale or at higher frequency resolution than what we solve for}.

{ 
\subsubsection{Number of fitted parameters vs. number of measurements}
\label{sec:dof}
An important consideration is to keep the number of fitted parameters small with respect to the number of measured visibilities to avoid flux loss and overfitting. We define the ratio between the number of measurements and fitted parameters as $q$. With $n$ stations we have $n(n-1)/2$ complex visibilities per (10~s) time slot per polarization. The solutions are obtained on data that is averaged down to a frequency resolution of 2~MHz. Considering the parallel hand polarizations (XX, and YY), we thus have 86,130 complex measurements in the 120-181~MHz band (29~blocks of 2~MHz with two polarizations and $n=55$).

We fit for $6 \times nm =$~22,110~``phase parameters'', where $m=67$ the number of directions. The factor of $6$ comes from the three phase and three TEC values we fit per antenna across the 120--181~MHz band. Thus, 86,130 complex measurements are fitted with  22,110 real-valued parameters\footnote{The full complex visibility function is minimized during the solve, not only the phase part.} resulting in $q \approx 7.8$.

For a 10~min solution interval (60~time slots of 10~s) per 2~MHz block, we also fit for $2\times nm$ complex gains. In this case $q \approx 24.2$.

Combining both calibrations together, we obtain $q\approx  7.7$. Thus, the number of free model parameters is still significantly smaller than the number of measurements. Ratio $q$ is almost completely determined by the TEC and Stokes~I phase corrections\footnote{We ignore the fact that for a couple of directions our solutions intervals are longer.}. 

The above calculation only provides a quick check, as it does not take into account the S/N per measurement, the accuracy of the input calibration model, and the precise functional form that is minimized. However, to first order our value of $q$ seems to be sufficient to prevent overfitting, see Section~\ref{sec:comdde}. 

In our calibration we have traded off the number of measurements  with S/N to speed up the calibration, by averaging a factor of 20 in frequency before obtaining the calibration solutions. Without this averaging, keeping a frequency resolution of 2~channels per subband,  this would have resulted in a much higher ratio of $q \approx 154$ (see Section~\ref{sec:future} for options to decrease the number of fitted parameters).


}

\subsection{Adding back the other facet sources and imaging}
\label{sec:addbackfacet}
After we obtain a set of DDE corrections for a facet, we add back all other sources in the facet and assume that these fainter sources can be corrected using the same solutions as for the ``center'' of the facet containing the bright source (group). The DDE calibration solutions are applied at the original 2~channels per subband frequency resolution. This allows the Stokes I + TEC corrections to be applied on a channel to channel basis\footnote{Recall that in the previous step (Sect.~\ref{sec:DDEselfcal}), the calibration was carried out on an averaged (phase-shifted) dataset with a frequency resolution of 2~MHz.}. After the corrections are applied the data is averaged by a factor of 5 in frequency and 3 in time. The amount of frequency averaging is less than described in Sect.~\ref{sec:DDEselfcal} to avoid bandwidth smearing, because the largest facets have sizes of several tens of arcmin. We then image the facet using MS-MFS clean in {\tt CASA} ({\tt nterms=2, robust=-0.25}) and W-projection (and multi-scale if needed). 

Optionally, the facet images can be made with the {\tt WSClean} imager \citep{2014MNRAS.444..606O}. Note that at this point we do not use the {\tt awimager} \citep{2013A&A...553A.105T}, which can correct for the time-varying LOFAR station beams across the FoV during imaging. There are two reasons for this, (1) MS-MFS {\tt nterms>1} imaging was not yet fully implemented and (2) {\tt awimager} has problems with imaging regions significantly beyond the HPBW of the station beam (the beam response drops close to zero in some regions here and the large beam corrections result in instabilities).

\subsection{Subtracting all the facet sources}
\label{sec:subsources}
{After the imaging of a facet is completed, we have obtained an updated sky model for the region of the sky covering that facet. This sky model is then subtracted from data (which consists of 2 channels per subband at 10~s time-resolution, see Section~\ref{sec:addbackfacet}), with the corresponding DDE solutions. The new output data that is created should now have the flux from the part of the sky that is covered by the facet correctly removed, while before the sources in the facet were only approximately removed as they were subtracted with the direction independent self-calibration solutions, as described in Section~\ref{sec:selfcal} and~\ref{sec:selfcalsub}. As a quality check, we re-image the residual data at low-resolution (2\arcmin) to verify that the sources in the facet were indeed correctly subtracted and that the magnitude of the residuals has decreased with respect to the direction independent self-calibration subtraction.}

After subtraction, we then proceed with the next direction, and start again with the process described in Sect.~\ref{sec:addbackDDE}. This whole process is repeated for all directions. We thus gradually build up a DDE corrected view of the sky, see Figure~\ref{fig:scheme} for the schematic overview of this process. {After each direction, the residual visibility dataset becomes ``emptier'' as more and more sources are subtracted with the DDE calibration solutions, instead of the direction independent self-calibration solutions.}

We note that the order in which the facets are treated goes roughly with the brightness of the source (group) that defines a facet.  Thus the ``worst'' facets (with the largest calibration errors in the medium resolution images) are treated first so that at a later point they do not influence the DDE calibration for the facets that have a fainter source (group). 

\begin{figure*}[th!]
\begin{center}
\includegraphics[angle =0, trim =0cm 0cm 0cm 0cm,width=0.95\textwidth]{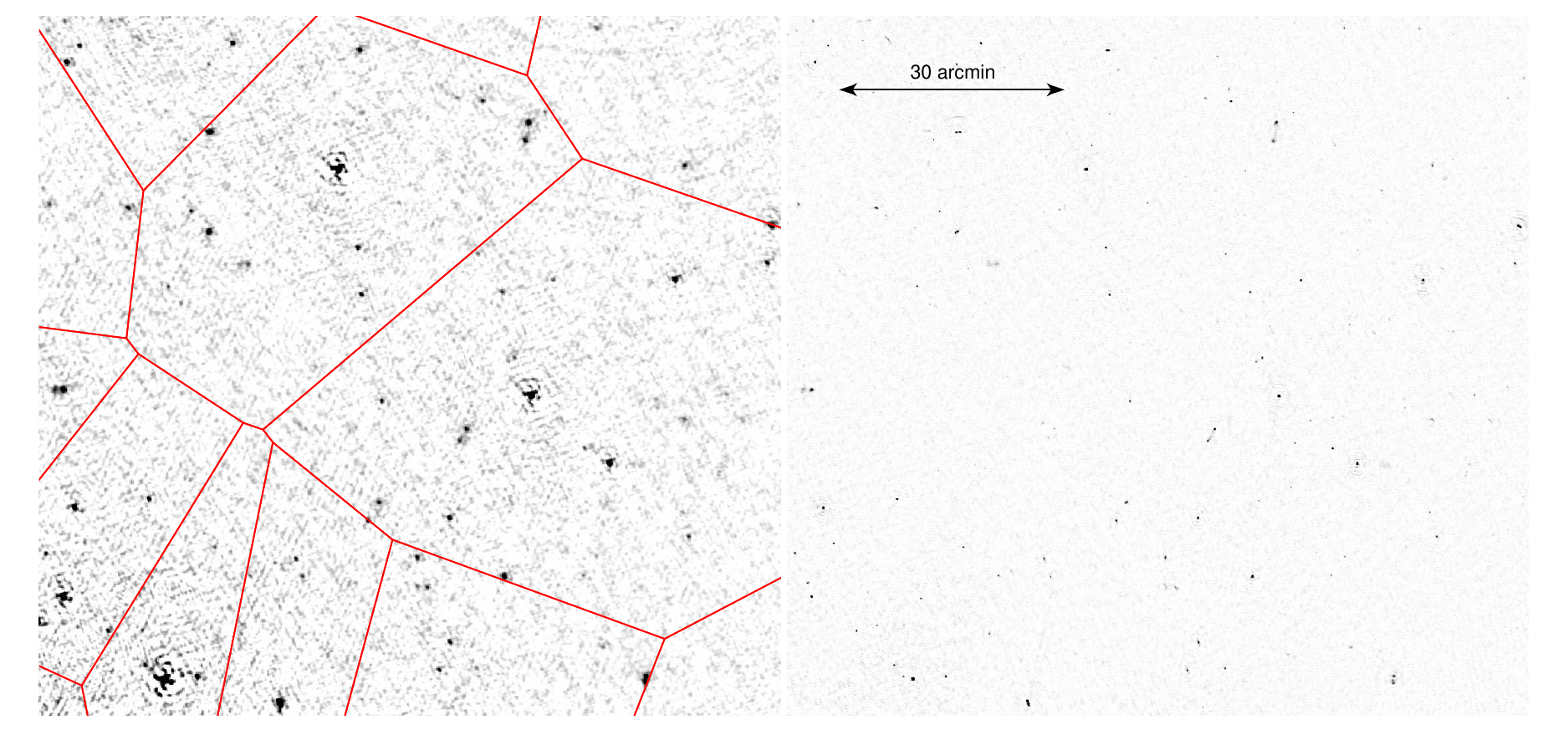}
\end{center}
\caption{{Comparison of a 25\arcsec~image at 150~MHz before facet calibration (left) and the high-resolution ($8.0\arcsec \times 6.5\arcsec$) full-bandwidth image after facet calibration (right). The r.m.s. noise level is close to 0.1~mJy~beam$^{-1}$ in the high-resolution image. The images cover the same part of the sky. Individual images of facets  were combined to make the high-resolution image. The outline of the facets is shown in red in the left panel. Images we manually scaled in brightness for the purpose of visual comparison.}}
\label{fig:postfactor}
\end{figure*}

{Depending on the science goals not all facets need to be calibrated, as the decrease in overall noise is mostly determined by the facets that contain the brightest ($\gtrsim 1$~Jy) sources (see also Section~\ref{sec:facetthoughts}).  In the end the individual facet images can be combined into an image that covers a larger (or the entire) FoV, which is mostly relevant for survey type science, see Figure~\ref{fig:postfactor}}.
This image is then corrected for the primary beam attenuation by dividing out the primary beam using an image of the beam obtained from {\tt awimager}.

{For survey type science, it is beneficial to re-image all of the facets after the facet calibration with the obtained DDE solutions. The reason for this is that the DDE calibration runs sequentially through the facets. Therefore, in particular the first facet that was produced did not have any other sources subtracted (from other facets) with the DDE calibration solutions. Such re-imaging is less relevant if the target of interest is located within the boundaries of a single facet and that facet is treated as the last one in the list, as was the case for the Toothbrush cluster. In \cite{williamsbootes} we will present a facet calibration run where re-imaging was carried out.}

In principle the whole DDE scheme could be iterated over; we did not attempt this due to computational limitations. We note though that subsequent iterations would offer much smaller improvements than in the first iteration, because the images from the first application of this scheme are already mostly free of calibration artifacts. 

\section{Discussion}
\label{sec:dicussions}

{ 

\subsection{Facet layout}
\label{sec:facetthoughts}
The amount of sky that needs to be imaged and corrected for direction dependent effects depends on the actual science case. We identify two main modes of operation for facet calibration (1) single targets of interest or (2) survey-type science. 

In the case of a single target, for example the observations described in this work,  it is actually not essential that the full FoV (within the HPBW) is imaged, or covered with facets. The removal of a limited number of bright sources already significantly improves the image quality for the target of interest. It was found that the  improvement in image noise after removal of the $>1$~Jy sources was relatively small, if the facets that were treated were located far away ($\gtrsim 1\degr$) from the target of interest.

For survey science, where the main aim is to image as much of the sky as possible, it is important to cover the entire region within the HPBW. If only a few facet directions are chosen, in combination with a maximum image size, only part of the field will be imaged and corrected by the facet calibration scheme which might not be desirable. Therefore, a sufficient number of calibrator sources are needed that are spread across the region within the HPBW. For example, with less than $\sim 20$~directions the assumption that calibration solutions are constant within a facet breaks down. From tests on about a dozen fields, we conclude that in general a sufficient number of calibrator sources is available (always more than 20) to achieve close to thermal noise limited images for a typical 8~hr synthesis run. Results for these fields will be described in upcoming papers. 

For fields at low declination (i.e., DEC~$\lesssim 10\degr$), it might be more challenging to find a sufficient number of calibrator sources to correct for direction dependent effects across the entire region within the HPBW of the station beam. This is caused by the reduced sensitivity at lower elevations. Therefore this requires a higher integrated flux density limit and thus reduces the number of calibrator sources available. It remains to be determined if this will become an important limitation to carry out facet calibration. Similarly, to achieve much lower noise levels (a few tens of $\mu$Jy~beam$^{-1}$, or less) higher dynamic range might be required and this could imply that the sky needs to be divided up into more facets. Due to the computational challenges involved in testing this (see Section~\ref{sec:compute}), we leave this as future work.

\subsection{Comparison with SPAM and Sagecal}
\label{sec:comparisonspamandsage}

Besides facet calibration, other direction dependent calibration schemes have been developed to deal with low-frequency radio data.    
We discuss some of the similarities and differences with the {\tt SPAM}  \citep{2009A&A...501.1185I} and {\tt Sagecal} \citep{2008arXiv0810.5751Y,2011MNRAS.414.1656K} packages. {\tt SPAM} has successfully been applied to a large number of GMRT and VLA datasets. {\tt Sagecal} has mainly been used for LOFAR HBA observation taken for the Epoch of Reionisation (EoR) KSP. In Table~\ref{tab:comparison} we provide a general comparison between these three different calibration schemes.

The main difference between facet calibration and {\tt Sagecal} are the calibration parameters that are solved for, the underlying solver, and the application of the calibration solutions. {\tt Sagecal}  obtains solutions for all directions at once, while facet calibration obtains solutions for a single direction at each step. {\tt Sagecal}  thus offers significant improvements in speed, also because the underlying solver is different from the ``traditional'' Levenberg-Marquardt solver that is used in the facet calibration. For more details about the solver employed by {\tt Sagecal} the readers is referred to the references provided above. 

{\tt Sagecal}  only solves for Jones parameters, so complex gain solutions are obtained for all four correlation products for each direction. The facet calibration takes a different approach, since it solves for phases on short timescales and parallel-hand complex gains on longer timescales. The main reason for this approach is that it enables a correction for the ionosphere on timescales as short as a few seconds, as phases can vary very quickly on the longer LOFAR baselines. Solving for a single phase component per direction per stations reduces the number of parameters that are solved for by a factor of 8 compared to full Jones. Solving for all Jones parameters on such short timescales is not feasible, because it would result in over-fitting due to the large number of d.o.f. Facet calibration also directly takes the $1/\nu$ ionospheric frequency dependence into account, further reducing the number of parameters that are solved for in the calibration. Another difference with respect to the facet calibration is that the data are not corrected with the direction dependent calibration solutions when imaged and deconvolved. In {\tt Sagecal}, sources are subtracted with the DDE solutions and optionally sources can be restored on an uncorrected residual image. 

It is important to note that {\tt Sagecal} was developed with a different goal in mind, namely the removal of sources from visibility data to detect the EoR, and not to obtain fully corrected images of the sky and correct for the ionosphere on short timescales. However, {\tt Sagecal} can subtract interfering  bright sources that would otherwise decrease the image quality of the main science target in the field, for example.

The facet calibration scheme is more similar to {\tt SPAM}  than {\tt Sagecal}, as {\tt SPAM}'s main goal is also to produce corrected images of the radio sky. 
{\tt SPAM} also attempts to solve for the ionosphere by obtaining phase solutions in a dozen or more directions on short timescales. {\tt SPAM} does currently not make use of the frequency dependence of the ionosphere, mainly because the GMRT and VLA bandwidths were too narrow to make this practical. A key difference with {\tt SPAM} is that the facet calibration scheme currently does not obtain a global phase screen from the phase solutions. Attempts to do this with using LOFAR data have only been partially successful till now. 
The production of a global phase screen to model the ionosphere above the array would  help to further reduce the number of d.o.f. in the calibration. Another difference with facet calibration is that {\tt SPAM} relies on {\tt AIPS} \citep{2003ASSL..285..109G} and is not well suited to handle LOFAR data. In addition, {\tt AIPS} does not allow one to solve for more general calibration problems. The underlying solver used in the facet calibration ({\tt BBS}) is more flexible, allowing for example to solve for TEC. In addition, in the facet calibration more emphasis is placed on solving for beam errors, while in {\tt SPAM} this is less relevant as it was not meant to work with phased arrays.

\begin{table*}
\begin{center}
\caption{Comparison Facet Calibration, SPAM, and Sagecal}
\begin{tabular}{llllll}
\hline
\hline
&Facet Calibration   & SPAM & Sagecal \\
\hline
Main purpose  & Corrected images the sky & Corrected images the sky  &  
Source subtraction\\
Solving for (scalar) phases & Y & Y & N\\
Solving for parallel hand gains  &Y &Y & Y \\
Solving for cross hand gains  &N$^{a}$  &N& Y \\
Correction and imaging of visibility data with DDE solutions &Y&Y&N$^{a}$\\
Explicit removal of instrumental effects (clocks) &Y&Y&N\\
Global phase screen modeling &N$^{b}$&Y&N\\
Optimized solver & N$^{c}$ & N & Y \\
Solutions obtained for all directions instantaneously & N & N &Y\\ 
Works on LOFAR HBA data & Y &N & Y \\
Solution intervals can vary per direction &Y&Y&Y \\
Solving for amplitude and phases on different timescales  &Y&Y&N \\
\hline
\hline
\end{tabular}
\label{tab:comparison}
\end{center}
$^{a}$ could be implemented and useful for polarization work\\
$^{b}$ being attempted\\
$^{c}$ {\tt Stefcal} \citep{2014A&A...571A..97S} can be employed for directions with $\gtrsim 1$~Jy of flux density\\
\end{table*}

\subsection{Flux-scale comparisons after DDE calibration}
\label{sec:comdde}
To test the effects of the facet calibration on the flux-scale we again compared the integrated LOFAR fluxes to those from the GMRT 150~MHz image (Section~\ref{sec:3C147bootstrap}). We corrected the LOFAR HBA image with the primary beam as described in Section~\ref{sec:subsources}. In addition, we corrected for the flux-scale difference found in Section~\ref{sec:selfcal}.

We extracted a source catalog from both images with {\tt PyBDSM}. For the GMRT map, a box size of $70$ pixels was used to compute the locally varying r.m.s. noise to take into account (calibration) artifacts around bright sources and variations due to the primary beam attenuation. For the HBA map we took a box size of $250$ pixels. We then cross-matched sources in the two catalogs using a matching radius of 2\arcsec. A S/N cut of 10 was imposed for all sources. Sources with an integrated flux over peak flux ratio larger than $2$ were excluded to avoid very extended and complicated sources. 

The resulting flux ratio of LOFAR over GMRT fluxes is shown in Figure~\ref{fig:gmrthbaflux}. We find that the overall flux-ratio between GMRT and LOFAR fluxes remains close to one. The medium of the flux-ratios is 1.02, which indicates the overall flux-scale was not significantly affected by the calibration. 

The overall spread in the flux density ratios is somewhat larger than what is expected from the error bars, but part of this extra scatter is related to the source extraction and the formation of sources from the individual Gaussians components (for more information regarding this the reader is referred to the {\tt PyBDSM} documentation). From the flux-ratio test we conclude that we do not find clear evidence for flux loss during the DDE calibration. {In addition, we recover faint sources and source structure in the LOFAR image \citep{vanweerenscience} that are also seen in higher frequency observations of the field \citep{2012A&A...546A.124V}, but are not included in our initial calibration model derived from the GMRT image at 150~MHz. The above findings are also supported by similar comparisons between images from LOFAR HBA facet calibration runs and GMRT images of the Bo\"otes \citep{williamsbootes} and H-ATLAS fields \citep{hardcastle}.
 However, we note that more detailed comparisons are needed to fully quantify the effects of the d.o.f. in our calibration. To do that, one can, for example, inject fake (faint) sources into the data and check how well they are recovered after facet calibration.} Given the processing requirements and amount of effort that will be required to carry out such tests, we leave this for future work.

\begin{figure}[th!]
\begin{center}
\includegraphics[angle =180, trim =0cm 0cm 0cm 0cm,width=0.45\textwidth]{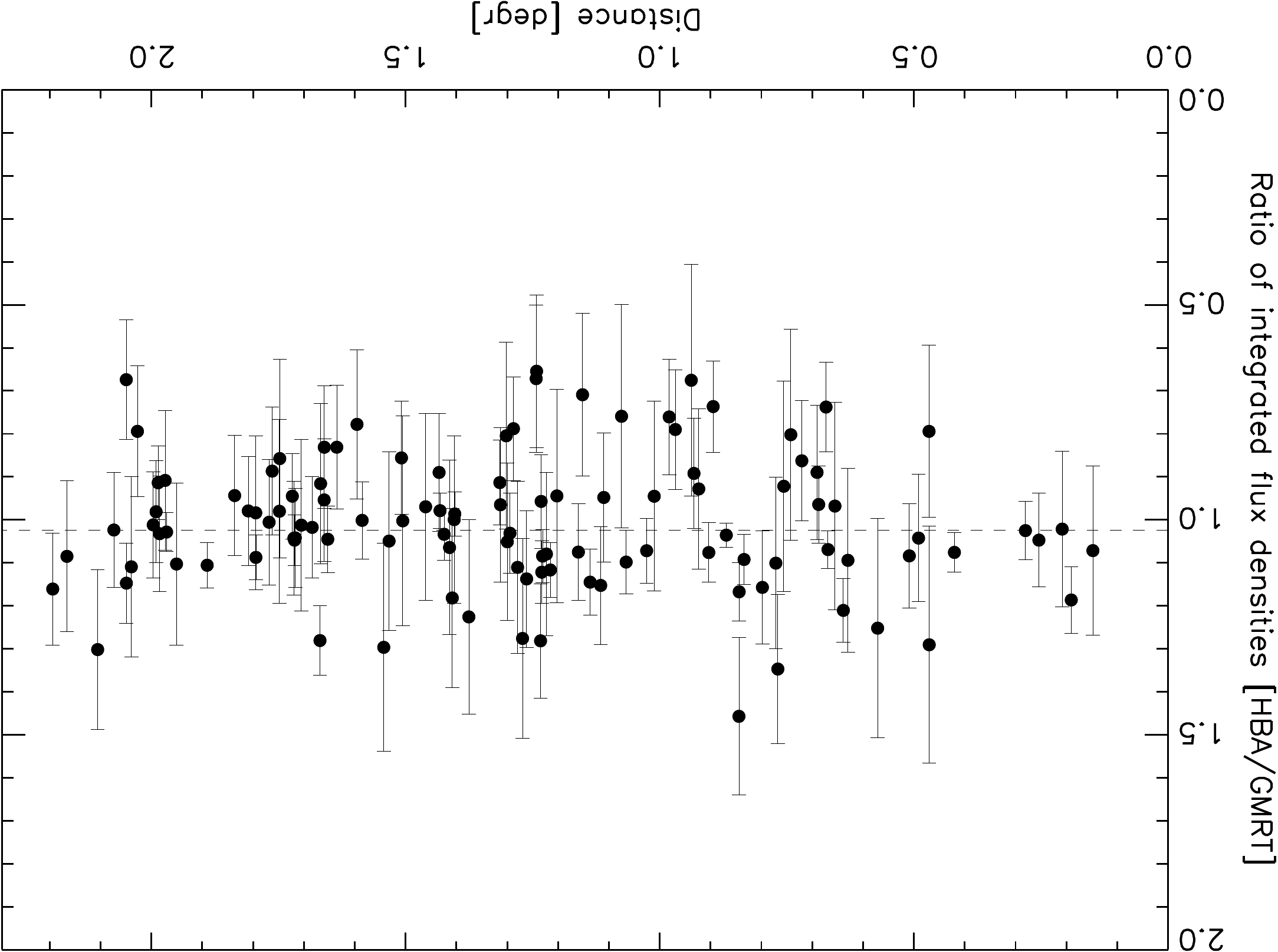}
\end{center}
\caption{{Flux density ratio between GMRT and HBA detected sources as a function of distance to the pointing center.}}
\label{fig:gmrthbaflux}
\end{figure}

\subsection{Computing requirements}
\label{sec:compute}

The amount of computing power required for facet calibration is considerable. For the Toothbrush field, we utilized a single 24 core (two 12-core Xeon E5-2695v2 CPUs) machine with 128~GB RAM. The total amount of pure processing time was 4 months. Therefore, if the facet calibration is to be applied to many more fields, it is  desirable to reduce the processing time. 

One option that has recently been exploited is to make use of {\tt WSClean} \citep{2014MNRAS.444..606O} instead of the {\tt CASA} imager. This typically reduces the imaging time  by a factor of 3 and decreases the overall processing time by about 30\%. In addition, significant speed-up has also been achieved by utilizing {\tt StefCal} for the calibration \citep{2014A&A...571A..97S}. Since {\tt StefCal} cannot solve for TEC at the moment, this is currently limited to the brighter ($\gtrsim1$~Jy) sources because more S/N is required as the bandwidth needs to be divided into smaller frequency chunks, where the effects of TEC can be neglected. A drawback is that this also increases the number of d.o.f. Another option to reduce processing time is to further parallelize the code. For example, once the brightest sources have been dealt  with (and subtracted) several different facets could be imaged and calibrated in parallel.

\subsection{Recent developments and future improvements}
\label{sec:future}
At the moment the facet calibration needs to be run in a semi-manual way by the user, which means that the method is currently more suited towards individual pointings rather than large surveys for which automated processing is required. To achieve more automated processing, several steps are required.

One is the automatic generation of the list of directions that define the facets. A starting point for the generation of such a list is a source list ordered by integrated flux density. Another aspect concerns the spatial extent of sources, because enough compact flux is needed to obtain calibration solutions for the distant remote LOFAR stations. Finally, a grouping algorithm is required to group closely separated sources. 

Another aspect concerns the situation in which facets boundaries can run across (extended) sources; this would preferably be avoided. The order in which the facets are calibrated in is currently also decided by the user, but this is relatively simple to automate. The order could follow the integrated (compact) source flux density. Furthermore, after the brightest sources are removed the order becomes less important.


Another improvement would be to increase the amount of S/N available for the DDE calibration. For example, it is possible that more facets are needed to reach noise levels of a few tens of $\mu$Jy~beam$^{-1}$ (i.e., Surveys KSP Tier-2 and 3 depth). For the purpose of correcting DDEs, more facets are better as corrections in more directions can be applied. On the other hand, depending on the source structure and flux density, as well as the quality of the subtraction from earlier processed facets, at some point the calibration solutions become too noisy for accurate subtraction. A way to further improve the S/N per direction and decrease the number of d.o.f., is to exploit the fact that solutions vary smoothly as a function of time. In addition, the ``slow gain solutions'' (correcting for the beam) should vary relatively smoothly across the HBA band. This information is currently not used in the facet calibration. {Calibration schemes that employ the spatial, frequency and/or time coherency of the calibration solutions have recently been developed by \cite{ 2014A&A...566A.127T,2015MNRAS.449.4506Y,2015MNRAS.449.2668S}}.

Work is ongoing on all above the above mentioned aspect. Ultimately, ionospheric phase screens and updated beam models (or amplitude screens) would provide an even larger reduction in the number of d.o.f., but this likely still requires a significant amount of work and study. The same holds for extending the facet calibration to the LBA. In the LBA, the ionospheric effects become more severe and differential Faraday Rotation cannot be neglected. A more fundamental limitation is the sensitivity of the LBA stations. LBA stations are about an order of magnitude less sensitive compared to the HBA and it is unclear if solutions in a sufficient number of directions ($\gtrsim 20$) can be obtained in the LBA frequency band.

\section{Conclusions}
\label{sec:conclusions}
In this paper we have presented a new calibration scheme to obtain deep high-resolution LOFAR HBA images. We applied this facet calibration scheme to the Toothbrush galaxy cluster field. The scientific  results are discussed in \cite{vanweerenscience}. 

This calibration scheme consists of a direction independent and a direction dependent part. For the direction independent calibration, the LOFAR clock offsets and flux-scale are determined by utilizing the gain solutions from a primary calibrator source. For the direction dependent calibration, the sky is divided up into facets, with each facet center being defined by a bright source or source group. Calibration solutions for the bright source (group) are obtained, solving for phases on short ($\sim 10$~s) timescales and parallel hand gains on longer timescales ($\sim 10$~min). The calibration solutions are applied under the assumption that they are constant across the facet. The updated model of the sky covered by the facet is then subtracted from the data with the solutions obtained. This scheme is repeated for subsequent facets, {slowly} building up a picture of the full FoV.

}
\acknowledgments
{\it Acknowledgments:}
We would like to thank the anonymous referee for useful comments. 
R.J.W. was supported by NASA through the Einstein Postdoctoral
grant number PF2-130104 awarded by the Chandra X-ray Center, which is
operated by the Smithsonian Astrophysical Observatory for NASA under
contract NAS8-03060. Support for this work was provided by the National Aeronautics and Space Administration through Chandra Award Number GO3-14138X issued by the Chandra X-ray Observatory Center, which is operated by the Smithsonian Astrophysical Observatory for and on behalf of the National Aeronautics Space Administration under contract NAS8-03060. G.A.O. acknowledges support by NASA through a Hubble Fellowship grant HST-HF2-51345.001-A awarded by the Space Telescope Science Institute, which is operated by the Association of Universities for Research in Astronomy, Incorporated, under NASA contract NAS5- 26555. G.B. acknowledges support from the Alexander von Humboldt Foundation. G.B. and R.C. acknowledge support from PRIN-INAF 2014. W.R.F., C.J., and F.A-S. acknowledge support from the Smithsonian Institution. F.A.-S. acknowledges support from {Chandra} grant G03-14131X. C.F. acknowledges support by the Agence Nationale pour la Recherche, MAGELLAN project, ANR-14-CE23-0004-01. Partial support for L.R. is provided by NSF Grant AST-1211595 to the University of Minnesota. 
Part of this work performed under the auspices of the U.S. DOE by LLNL under Contract DE-AC52-07NA27344.
LOFAR, the Low Frequency Array designed and constructed by ASTRON, has facilities in several countries, that are owned by various parties (each with their own funding sources), and that are collectively operated by the International LOFAR Telescope (ILT) foundation under a joint scientific policy. We thank the staff of the GMRT that made these observations possible. GMRT is run by the National Centre for Radio Astrophysics of the Tata Institute of Fundamental Research. The Open University is incorporated by Royal Charter (RC 000391), an exempt charity in England \& Wales and a charity registered in Scotland (SC 038302). The Open University is authorized and regulated by the Financial Conduct Authority.

\bibliography{ref_filaments}
\end{document}